\pdfoutput=1 

\documentclass{icrc2009}

\usepackage{graphicx}   
\usepackage[caption=false]{caption}    
\usepackage[font=footnotesize]{subfig} 
\usepackage{fixltx2e}
\usepackage{amssymb}
\usepackage{url}

\newcommand{\shorttitle}[1]%
{\markboth{Proceedings of the 31\MakeLowercase{$^{st}$} ICRC, {\L}\'{o}d\'{z} 2009}{#1} }
\newcommand{\etal}{\MakeLowercase{\textit{et al. }}} 

\begin{document}
\title{Rapporteur Summary of Sessions HE 2.2-2.4 and OG 2.5-2.7}

\author{\IEEEauthorblockN{Teresa Montaruli\IEEEauthorrefmark{1}}
                            \\
\IEEEauthorblockA{\IEEEauthorrefmark{1}University of Wisconsin - Madison, 1150 University Ave, 53706, USA}}

\shorttitle{T. Montaruli  Rapporteur Summary of HE 2.2-2.4, OG 2.5-2.7}
\maketitle

\begin{abstract}
The physics items presented in the HE 2.2-2.5  sessions  include a variety of 
results concerning neutrino oscillations, dark matter and anti-matter, supernova neutrinos and proton decay. The OG 2.5-2.7 sessions concern detector R\&D and operations in gamma and neutrino astronomy.
I report here about a selection of the presented results. These indicate that Neutrino Astronomy is in a mature phase. Challenging detectors are being operated and produce results of interest. Their sensitivity is now getting closer and closer to the region of interest for various predictions. For galactic sources, if expected neutrino fluxes are derived from measured gamma fluxes, it is possible that a few years of operation of cubic-km detectors are needed to produce a statistically significant discovery. The era of the cubic-km detectors is now a reality in the Antarctic ice. The Mediterranean sea and Lake Baikal are hosting smaller scale detectors and proposing future extensions to the cubic-kilometer scale. 
Long baseline experiments and Super-Kamiokande are measuring with increasing precision the neutrino mixing matrix elements while waiting for dedicated experiments to measure $\theta_{13}$. Indirect detection of dark matter (DM) with gammas and neutrinos covers a complementary parameter space than LHC and direct detection experiments. Even if these measurements are affected by astrophysical uncertainties, they can provide a hint on the nature of the dark matter particle. Anti-matter limits on deuterons and anti-protons from PAMELA and balloon experiments, such as BESS-Polar, are of high interest and AMS-02 will extend the sensitivity to higher energy and lower fluxes.
While past conferences (such as ICRC2003) where dominated by MeV-GeV neutrino results on atmospheric and solar oscillations, this conference was dominated by gamma astronomy results. The field is rich of R\&D programs that are especially intriguing for what concerns photo-detectors, such is the case of Geiger-APDs. Several works showed that there is room to increase sensitivities of Imaging Atmospheric Cherenkov detectors by not negligible factors with more clever analysis tools and topological triggers.
 \end{abstract}

 
\section{Introduction}
\label{sec1}
The sessions I have to summarize concern: HE 2.2 Observations on Solar and Atmospheric Neutrinos (13 contributions); HE 2.3 Search for new particles and phenomena (as dark matter,...) (30 contributions); HE 2.4 New Experiments and Instrumentation (25 contributions); OG 2.5 High Energy Neutrino Astrophysics (42 contributions); OG 2.6 Gravitational waves and Experiments (no contributions); OG 2.7 New Experiments and Instrumentation (83 contributions).  Most of the works are about experimental results of detectors in operation and a lot of R\&D studies especially for Imaging Cherenkov Atmospheric detectors. There were only a few theory contributions. The gravitational wave (GW) community was absent at this conference. It is possible that this is a sign that there is not yet enough exchange between the GW and astroparticle communities. This coming decade may be dominated by the effort of opening a new window on the universe using the yet unobserved GWs. Their detection will test General Relativity and help understanding astrophysical sources. The next decade may as well see the discovery of the other longest range messengers, astrophysical neutrinos. Moreover, ideally the discovery of neutrino and gravitational sources should be accompanied by electro-magnetic observations. The interconnections between GW, neutrino and gamma astronomy are so strong that multi-wavelength approaches are essential to provide the most complete possible picture of astrophysical sources and scientific communities working in these fields should hopefully increase their level of interaction. 

It is very difficult for me to find a guiding line between all of these different subjects to review. They share one aspect in my view: they address the question on the mass/energy content of the universe.
This paper is structured as follows: in Sec.~\ref{sec2}  I will address the question: what are we still learning on the problem of neutrino mass through oscillation measurements? In Sec.~\ref{sec3}  I will discuss upper limits on supernova collapse and proton decay. In Sec.~\ref{sec4} I will  
describe what was presented about the mass content in the Universe and its nature, specifically about  indirect searches for DM using gammas and neutrinos and on anti-matter. 
In Sec.~\ref{sec5} I will focus on searches for astrophysical sources in neutrino telescopes  focusing on point-like sources and diffuse fluxes. 
In Sec.~\ref{sec6} I will describe some of the R\&D programs and proposed experiments in the neutrino  and gamma astronomy fields.

\section{Neutrino oscillations in the atmospheric neutrino sector}
\label{sec2}

Currently, neutrino oscillation data can be accommodated in a framework where flavor and mass neutrino states are connected by a unitary mixing matrix that depends on 3 angles ($\theta_{12}$, $\theta_{13}$, $\theta_{23}$), one CP-violating phase $\delta$, and 2 squared mass differences, $\delta m^2 = m_{2}^2 - m_1^2$ and $\Delta m^2 = m_3^2 - (m_1^2 + m_2^2)/2$. While ($\delta m^2$,
sin$^2 2\theta_{12}$) and ($\Delta m^2$, sin$^2 2\theta_{23}$) are relatively well measured with solar and atmospheric neutrinos, only upper bounds have been set until now on $\theta_{13}$. CHOOZ found no evidence for $\bar{\nu}_e$ disappearance and set an upper bound in the range sin$^2 2\theta_{13} < $few \% \cite{chooz}  (CHOOZ limit at MINOS best fit value of $|\Delta m^{2}_{32}| = (2.43 \pm 0.13) \times 10^{-3}$ eV$^2$ is $sin^2 2\theta_{13} < 0.15$). Future investigations on leptonic CP violation and on mass hierarchy (i.e. sign of $\Delta m^2$) depend on a not null value of $\theta_{13}$. It has been shown in Ref.~\cite{fogli} that the combination of solar and atmospheric neutrino results with long-baseline reactor neutrino data implies  a non-zero $\theta_{13}$ at the level of $1\sigma$ and a preferred value of sin$^2 2\theta_{13} \sim 0.02 \pm 0.01$.
The recent MINOS $\nu_{\mu} \rightarrow \nu_{e}$ appearance results tend to push the value of this angle to a non-zero value.

\begin{figure*}[!t]
   \centerline{\subfloat[From top to bottom: allowed oscillation regions at 90\% (solid lines) and 68\% c.l.(dotted lines) by K2K (larger black contours), MINOS (green regions, best fit indicated by a square), SK L/E analysis (the blue star indicates the best fit $|\Delta m^2_{23}|$, sin$^2 2\theta_{23} = 2.2 \times 10^{-3}$ eV$^2, 1.04$) and SK zenith distributions (the red dot is the best fit $2.1 \times 10^{-3}, 1.01$).]{\includegraphics[width=2.5in]{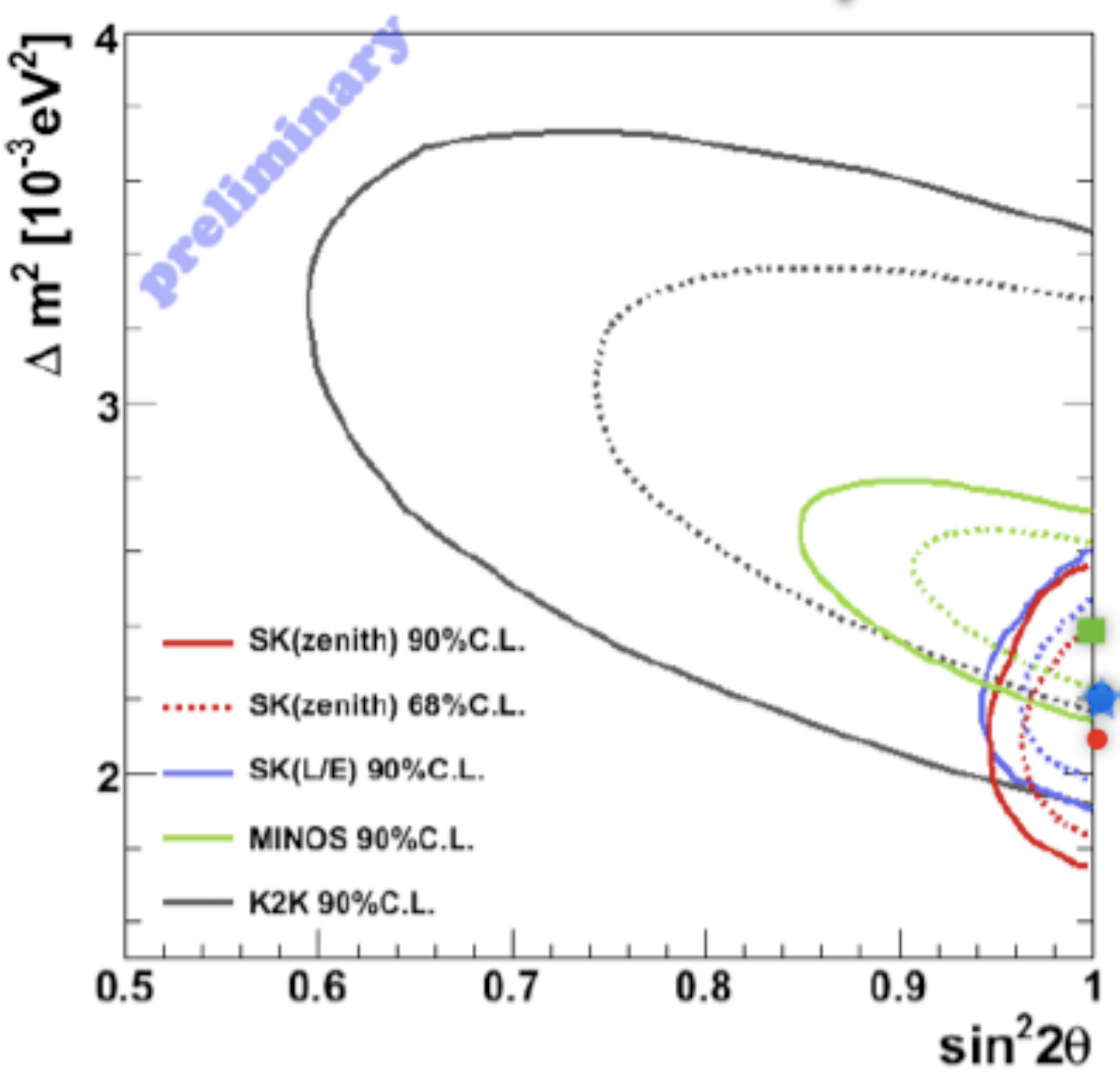} \label{left_fig0}}
              \hfil
              \subfloat[Range of values of sin$^2 2\theta_{13}$ and CP-violation phase $\delta_{\rm CP}$ that can produce the number of events selected in MINOS by an Artificial Neural Network (ANN) technique. The arrows indicate the allowed regions (90\% c.l. for normal and inverted hierarchies and the excluded region by CHOOZ \protect\cite{chooz}. Close to this limit the 2 best fit curves are also indicated for normal (solid line) and inverse (dotted line) hierarchies.]{\includegraphics[width=2.8in,height=2.5in]{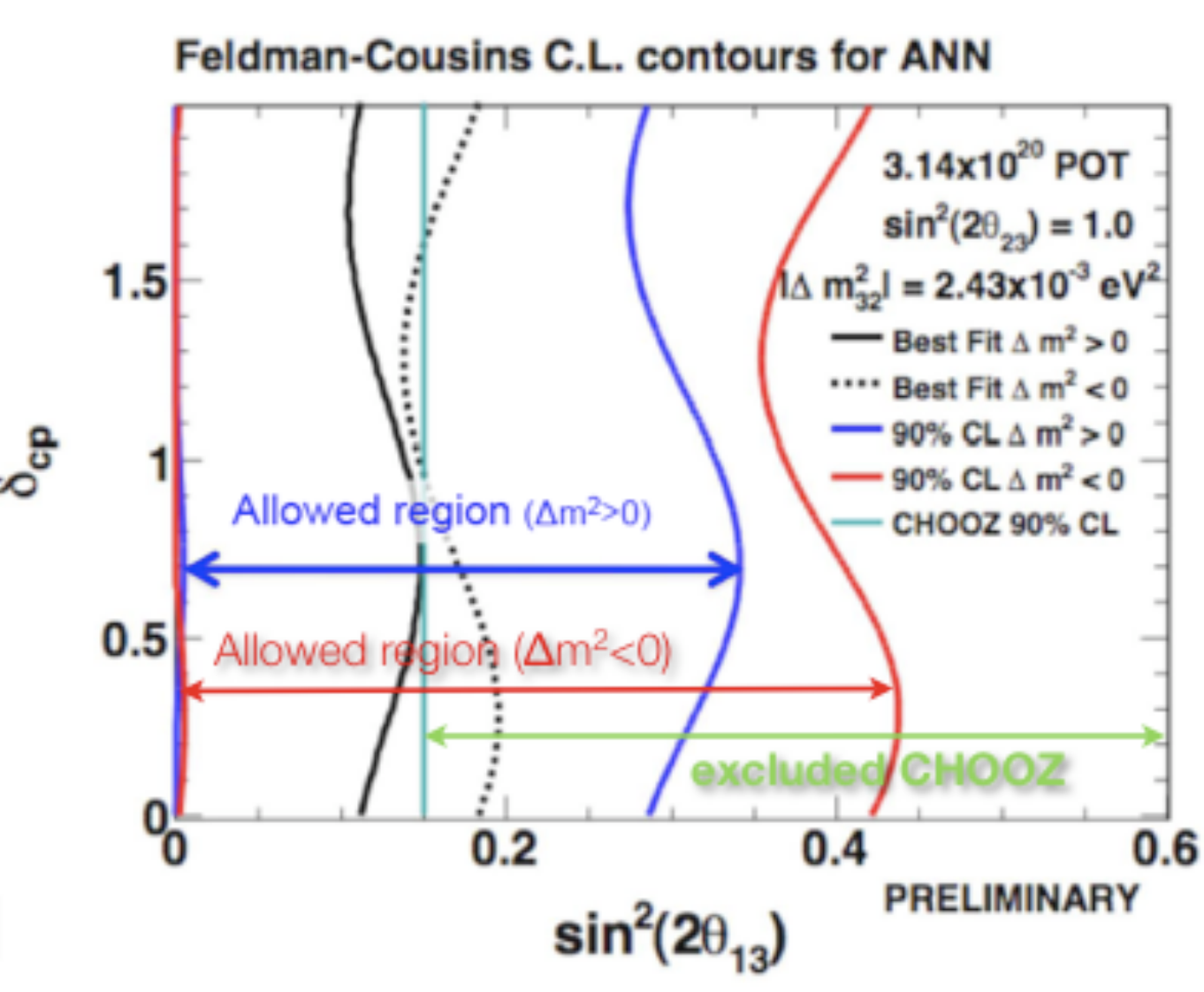} \label{right_fig0}}
             }
   \caption{}
   \label{fig0}
 \end{figure*}

MINOS uses an intense neutrino beam produced by 120 GeV protons accelerated by the Main Injector at Fermilab. At the time of the conference $7.6 \times 10^{20}$ protons on target (pot) were delivered to MINOS. The beam is intercepted by a 1 kt near detector (ND) at Fermilab and a 5.4 kt far detector (FD) at 735 km in the Soudan Underground laboratory. These sampling and tracking calorimeters have similar design with thick steel planes and polystyrene scintillators read by wavelength shifting fibers coupled with multi-anode photomultipliers (PMTs).
At the conference MINOS presented a summary of all results on oscillations \cite{MINOS}. Most of them were obtained in the low energy configuration with the target inserted in the first magnetic horn to maximize the number of neutrinos in the 1-3 GeV range where oscillation effects are important.
The beam contains 91.7\% $\nu_{\mu}$, 7\% $\bar{\nu}_{\mu}$ and 1.3\% $\nu_{e} + \bar{\nu}_e$.
The analysis cuts resulted in 848 $\nu_{\mu}$ charge current (CC) candidates in the FD compared to the unoscillated expectation of $1065 \pm 60_{\rm sys}$. The ratio of the events to the unoscillated prediction as a function of the reconstructed neutrino energy shows a dip
as expected from the disappearance probability function: $P(\nu_{\mu} \rightarrow \nu_{\tau}) =$ sin$^2 2\theta sin^2(1.27 \Delta m^2 L/E)$, where $L$ is the distance between the 2 detectors and $E$ the neutrino energy. 
The depth of the dip is connected to the mixing angle measurement and its position in energy is connected to $\Delta m^2$ in the survival probability. MINOS best fit values 
are $|\Delta m^{2}_{32}| = (2.43 \pm 0.13) \times 10^{-3}$ eV$^2$ and sin$^2 2\theta_{23} > 0.90$ at 90\% c.l. \cite{disappearance}. Alternative scenarios to oscillations such as neutrino decay and decoherence are disfavored at 3.7 and 5.7 standard deviations, respectively. 
The preferred regions in the parameter space of 2 flavor $\nu_{\mu}\rightarrow \nu_{\tau}$ oscillations for MINOS CC disappearance analysis, Super-Kamiokande (SK) and K2K are shown in Fig.~\ref{left_fig0}. SK-1+2+3 data were analyzed (2805.9 d for fully contained and partially contained events, 3109.2 d for upward going muons) and the zenith angle distribution for the various topologies characterized by different energy ranges were derived. Also the L/E analysis was updated. This analysis demonstrates the oscillatory behavior using the partially contained muon events and the fully contained events \cite{SK}. SK also did not find any deviation of $\theta_{23}$ from $\pi/4$ by looking at the low energy ($\sim 1$ GeV) electron data and derived the 90\%c.l. region $0.410 < $sin$^2 \theta_{23} < 0.585$.

MINOS analyzed the $\bar{\nu}_{\mu}$ oscillations to investigate CPT conservation. Between $3.2 \times 10^{20}$ pot, 42 events where selected in the FD while $64.6 \pm 8.0_{\rm stat} \pm 3.9_{\rm sys}$ are expected for no oscillations and $58.3 \pm 7.6_{\rm stat} \pm 3.6_{\rm sys}$ for CPT conserving oscillations at the best fit of the $\nu_{\mu}$ disappearance analysis. The statistics is still inconclusive and results are compatible with oscillations under CPT conservation. The event statistics is as well limited for the appearance of electron neutrino analysis that selected 35 events that are compatible at $1.5\sigma$ level with $27 \pm 5_{stat} \pm 2_{\rm sys}$ expected background events. Evidence for electron neutrino appearance would imply a non-zero $\theta_{13}$ and results depend also on the CP-violation phase $\delta$, on the mass hierarchy and other oscillation parameters. At this stage, the experiment indicates the allowed regions in Fig.~\ref{right_fig0}. In the future the NO$\nu$A experiment
\cite{NOVA} will push limits from the region of about 10\% down by about an order of magnitude. SK limits for direct (inverted) hierarchy are $sin^2 \theta_{13} < 0.075 (0.13)$  \cite{SK}. T2K run is also going to start at the end of the year.

The SNO experiment \cite{SNO} presented the atmospheric neutrino induced events angular distribution and used it for deriving an allowed region for 2-flavor oscillations. Even if the statistics of the experiment is limited (514 events in 1229.3 d), the allowed region agrees with the current scenario. SNO best fit is $2.6 \times 10^{-3}$ eV$^2$ and maximal mixing for a flux normalization with respect to the Bartol flux \cite{bartol} of about 1.22 (the spectral index was not fit).
Being the experiment extremely deep (5890 mwe) atmospheric muons can be rejected up to cos$\theta = 0.4$ above the horizon and 201 atmospheric neutrino events were identified in this region.  

\subsection{Atmospheric Neutrinos and Muons}
\label{sub-neu}

\begin{figure*}[!t]
   \centerline{\subfloat [Unfolded muon neutrino spectrum measured with 22 strings of IceCube \protect\cite{dima_ic22}, averaged over zenith angles larger than $90^{\circ}$ compared to simulation using the Bartol flux \protect\cite{bartol} and to AMANDA-II results in \protect\cite{atmnu_amanda} and in \protect\cite{kirsten}. The grey band is the $1\sigma$ statistical error of the unfolding and it takes into account the difference between the unfolded spectra using data with the centre of gravity of hits in the bottom and top of the detector. This difference is due to the fact that IceCube simulation does not perfectly account for the depth dependency with ice.  
The collaboration is devoting significant efforts to understand and reduce systematic 
uncertainties as the statistics increases. The red histogram represent the spectrum from simulation using the Bartol flux \protect\cite{bartol}.]{\includegraphics[height=2.8in,width=2.8in]{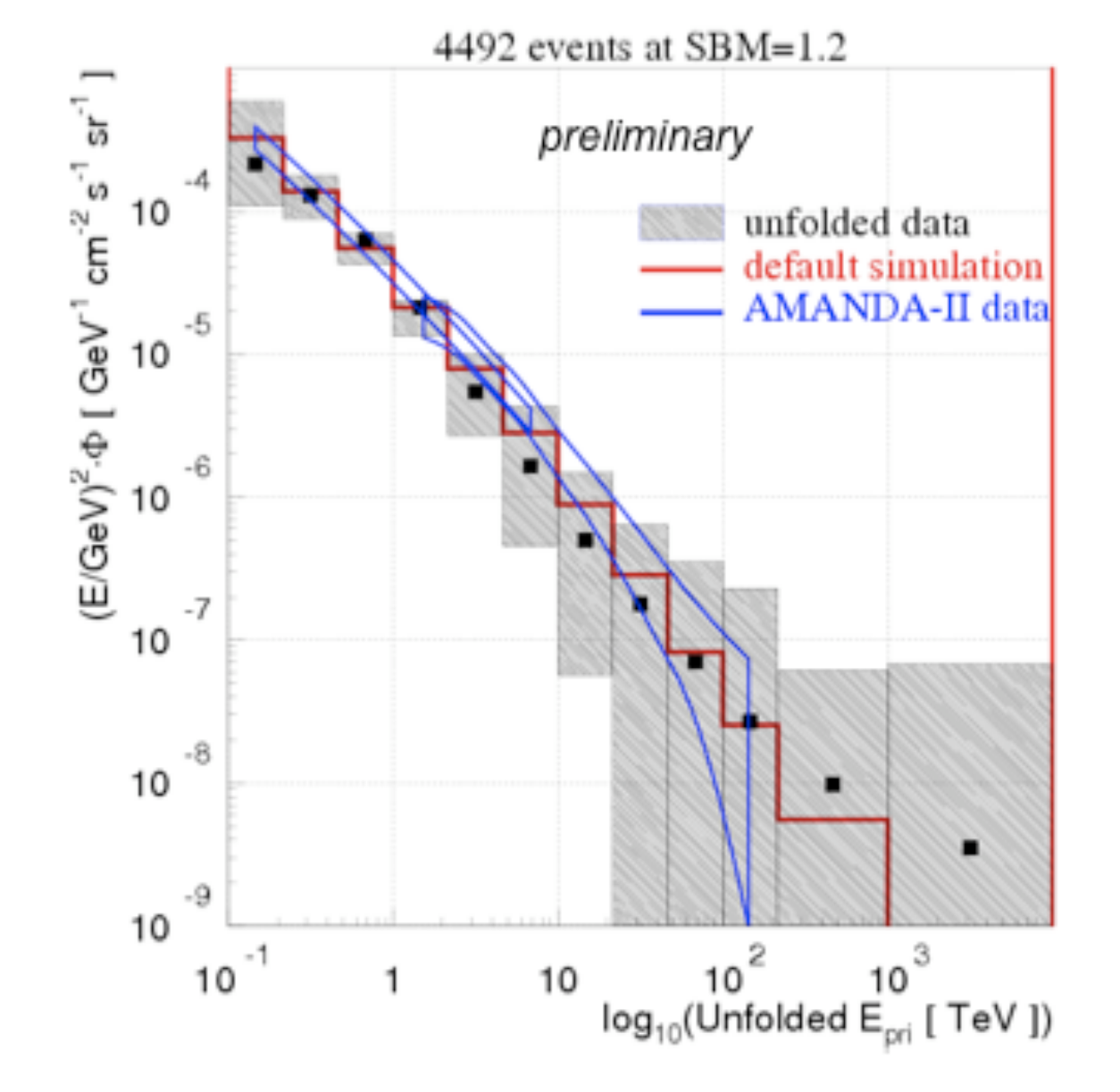} \label{left_fig1}}
              \hfil
              \subfloat[IC22 atmospheric neutrino spectrum compared with predictions for $\nu_{\mu} + \bar{\nu}_{\mu}/3$. The factor $1/3$ accounts for the lower CC interaction cross section of anti-neutrinos up to about $10^{5}$ GeV compared to neutrinos for neutrino induced muon events in IceCube. Conventional neutrinos are calculated as in \protect\cite{honda}, prompt models are calculated in the framework of perturbative-QCD in Refs.~\protect\cite{sarcevic} (std is for optimal parameters, min and max indicate the range of variation of parameters) and in \cite{martin}  for different structure functions. Also the RQPM model in \protect\cite{naumov} is shown.]{\includegraphics[height=2.6in,width=2.8in]{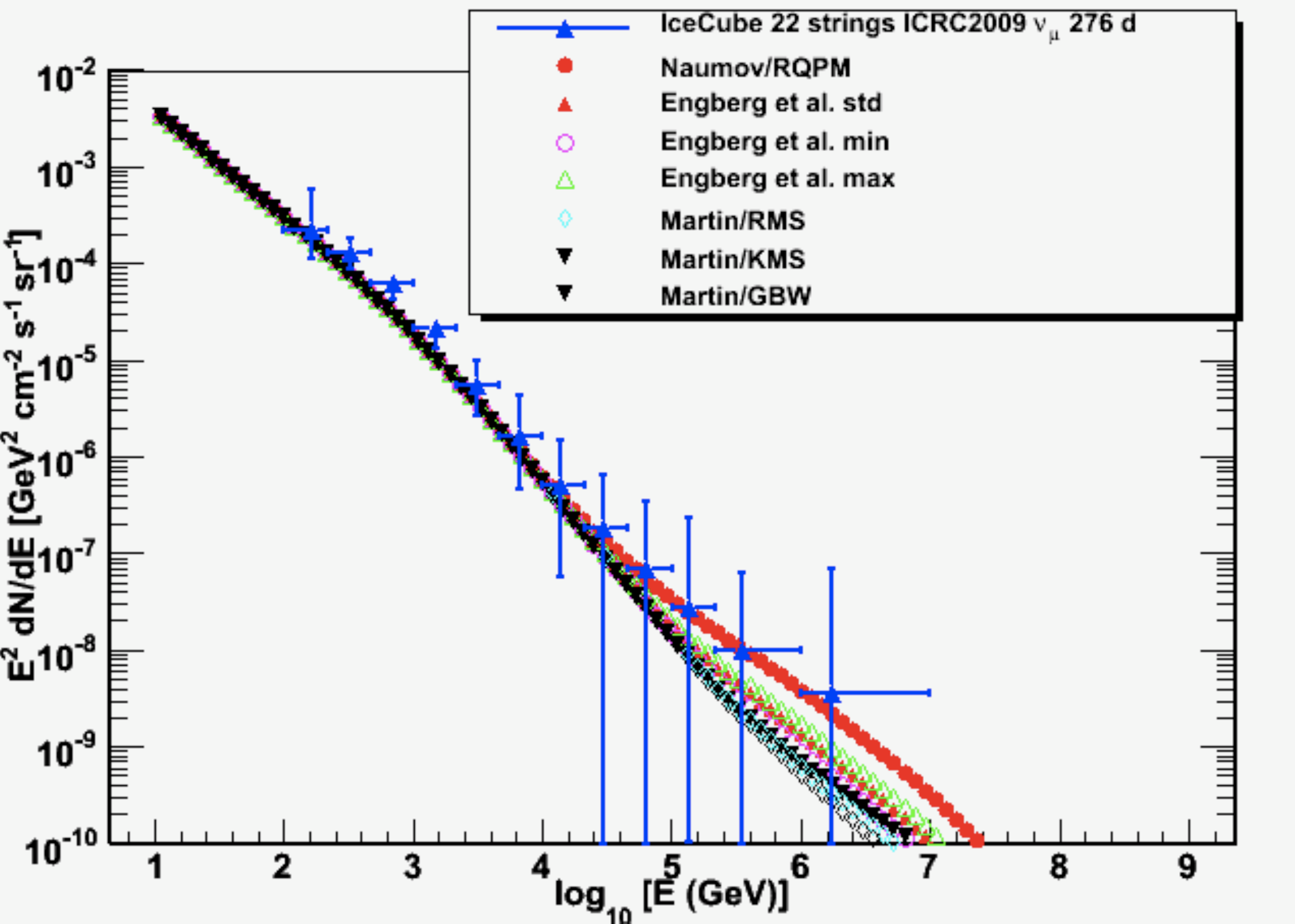}\label{right_fig1}}
             }
   \caption{}
   \label{fig1}
 \end{figure*}

At higher energies (E $\gtrsim 100$ GeV and up to about 10 TeV), where oscillations do not affect anymore the atmospheric neutrino beam, a two-parameter (normalization and spectral index) forward folding of the atmospheric neutrino flux was recently performed by AMANDA-II to determine variations with respect to existing models \cite{atmnu_amanda}. An allowed region in the parameter space normalization vs spectral index was derived relative to the Bartol flux using the angular distribution and an energy proxy for 5511 neutrino events collected in 1387 d. The best fit point indicates that the data prefer a higher normalization by about 10\% and a slightly harder spectrum by about 0.056 at 640 GeV. At this conference IceCube presented an atmospheric neutrino measurement with the 22 string configuration \cite{dima_ic22} that spans a wider energy range up to PeV energies. This is compared to the AMANDA-II measurement in Fig.~\ref{left_fig1}. 
A sample of 4492 events collected in 275.5 d with a contamination of 5\% from misreconstructed atmospheric muons was selected. The muon energy was reconstructed based on the fact that the energy loss is proportional to the muon energy when stochastic energy losses due to bremsstrahlung, pair production and photonuclear interactions  dominate compared to 
the continuous ionization losses. The muon energy reconstruction has a resolution of about 0.3 in log$_{10} E_{\mu}$ between $10^4 \div 10^8$ GeV. An unfolding method was applied to extract the neutrino energy spectrum from the atmospheric muon neutrino data. The main systematic error that affects this measurement is the depth dependence of the ice which is still not perfectly simulated.
The error is so large at high energy that it is not yet possible to disentangle at energies $> 10$ TeV an eventual contribution of prompt neutrinos from decays of charmed mesons and baryons or an astrophysical component \cite{sarcevic,sine}, as it is visible in Fig.~\ref{right_fig1} where various models for prompt neutrinos are shown with the measurement. Neutrinos from $\pi$s and $K$s, indicated as conventional neutrinos, have a differential energy spectrum above 100 GeV of about $E^{-3.6}$ and higher flux at the horizon because decay of mesons is advantaged with respect to the vertical. On the other hand, prompt neutrinos are characterized by a harder spectrum that resembles the one of primaries (before the knee $\sim E^{-2.7}$, after about $E^{-3.1}$). Moreover, since charmed mesons have very short decay lengths, the angular distribution is flat in zenith. Neutrinos directly produced in sources would instead follow an $E^{-2}$ power law characteristic of $1^{st}$ order Fermi mechanisms. 
Uncertainty in calculations of atmospheric neutrinos above 10 TeV are a problem for neutrino telescopes since prompt neutrinos are a hard component that contaminates the signal region. Disentangling the two components will be challenging in diffuse flux analyses from extra-galactic sources \cite{ic22_kotoyo}.

High energy atmospheric neutrinos in neutrino telescopes such as IceCube are dominantly produced by 
kaons. On the other hand, kaon production starts to be dominant for atmospheric muons  at much higher energies. As a matter of fact, even if they come from the same decay as neutrinos, the energy fraction taken by the muon is much larger than that the neutrino. At 1 TeV more than 75\% of neutrinos (depending on the hadronic model) are produced in kaon decays and the rest in pion ones, while only 20\% of the muons come from kaons.
Models that foresee a larger $K^+$ production (such as SIBYLL 2.1) show better agreement than others
(e.g. QGSJET II) \cite{sine,neutrino2008}. The preference for such hadronic models is also indicated by the muon ratio measurement of MINOS \cite{MINOS_muons} that shows a rise between 0.3-1 TeV consistent with an increasing contribution to the muon charge ratio from kaons. Consistent results were obtained by another deeper magnetic spectrometer, OPERA, that intercepts the CNGS neutrino beam from CERN at the national laboratory of Gran Sasso \cite{opera}.
While OPERA is dedicated to $\nu_{\tau}$ appearance discovery, at this conference they proved their
cosmic ray capability by measuring the muon charge ratio for single and multiple muons. They also derived the primary cosmic ray spectrum from the secondary muon one, and showed preliminary results from a study of large $p_{T}$ events (e.g. from charm) in coincidence with the LVD scintillator detector. Such study is also ongoing in IceCube \cite{gerhardt}.
The experiment LHCf  \cite{LHCf}, dedicated to the measurement of neutral particles, such as neutrons and $\pi^0$, emitted in the very forward region (of interest for cosmic rays) and TOTEM at LHC will improve the understanding of hadronic models at laboratory energies of about $10^{17}$ eV.
 
\begin{figure*}[!t]
   \centerline{
             \subfloat[Elevation $ = \pi/2$ - zenith distribution for atmospheric muons and atmospheric neutrinos for ANTARES multi-line data collected in the 9-12 line configuration  (black histogram with statistical errors) compared to simulation \protect\cite{aart}. There are 582 uward-going candidate neutrino events while the simulation predicts 494 atmospheric neutrinos and 13 mis-reconstructed muons. The shaded band indicates the error of the simulation.]{\includegraphics[height=2.8in,width=2.8in]{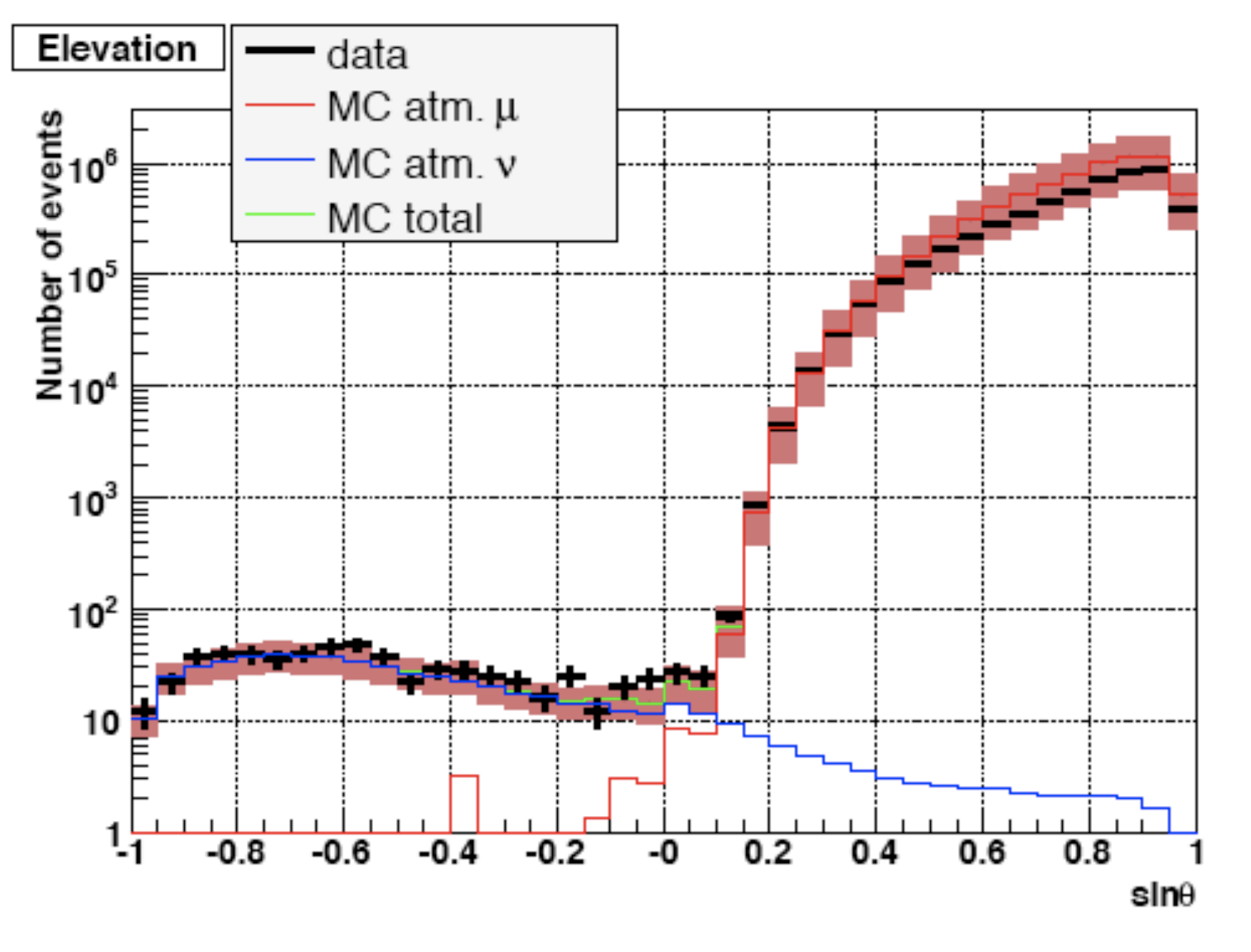}\label{left_fig2}}
                         \hfil
                 \subfloat[Atmospheric muon vertical intensity measured by under-water and ice arrays as a function of depth compared to calculations (all references in in Ref.~\protect\cite{bazzotti}). MUPAGE is a fast simulation from a parametrization of deep muon data ($E_{\mu} > 20$~GeV) in the MACRO underground detector \protect\cite{MUPAGE}.]{\includegraphics[height=2.8in,width=3.in]{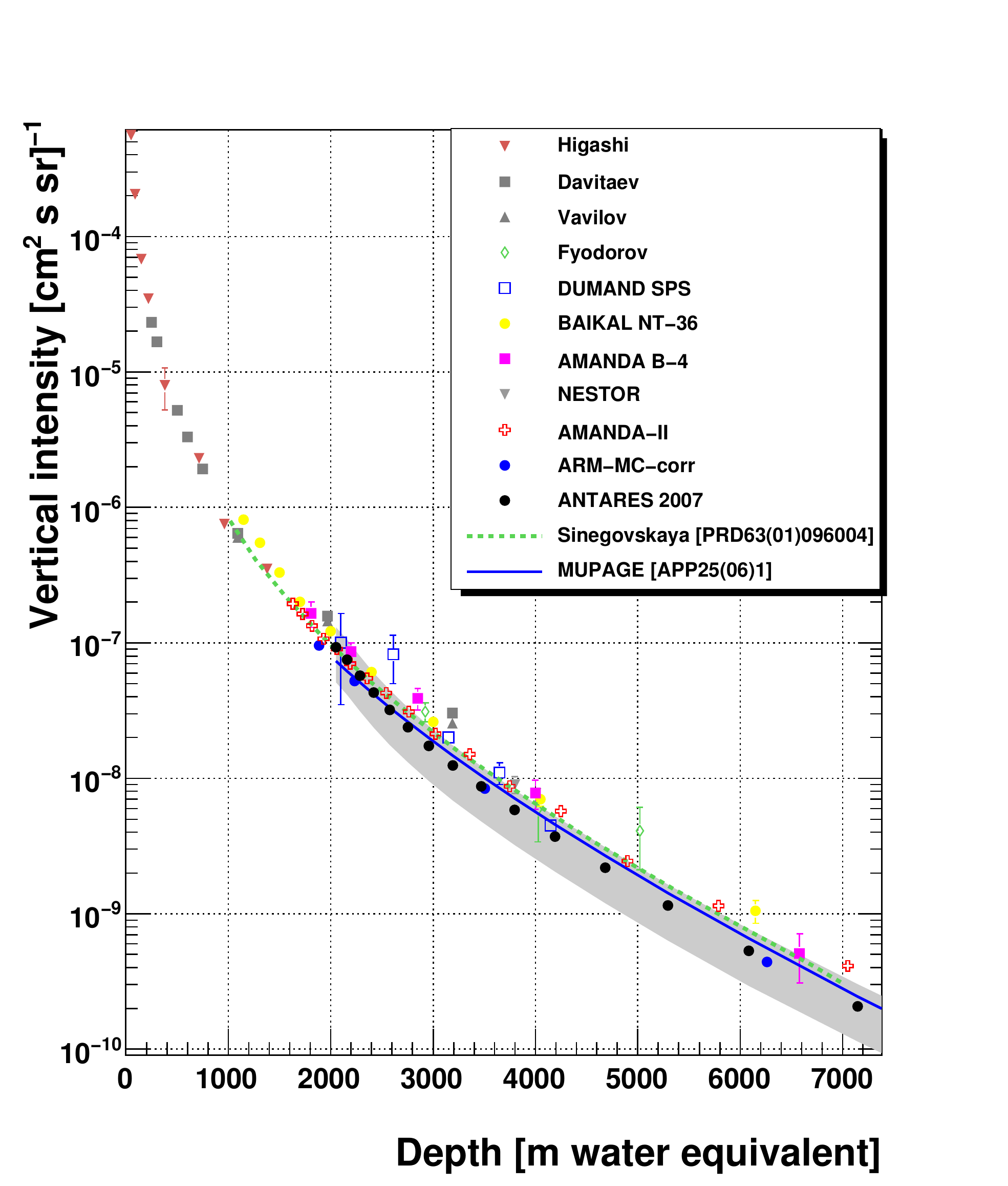}\label{right_fig2}}
            }
   \caption{ }
   \label{fig2}
 \end{figure*}

Deep neutrino telescopes, such as IceCube \cite{karle}, ANTARES \cite{aart} and SNO \cite{SNO}, proved their ability to see the transition between atmospheric muons and neutrinos in their zenith angular distribution. This is an important test that verifies the tracking capability and the detector understanding through comparison with simulations of atmospheric neutrinos and muons. As an example the ANTARES all-sky zenith distribution for data and MC is shown in Fig.~\ref{left_fig2}.
The vertical intensity of muons as a function of depth for deep detectors was presented in in Ref.~\cite{bazzotti,SNO}. In Ref.~\cite{bazzotti} a collection of underwater detectors and prototype muon vertical intensities (see Fig.~\ref{right_fig2}) was presented and compared to a fast simulation derived from a parametrization based on MACRO data \cite{MUPAGE}.

\section{Supernova neutrinos and nucleon decay}
\label{sec3}

There are two different kinds of supernova explosions, type Ia and core collapse supernovae (type Ib, Ic and II). Type Ia are believed to occur when a carbon-oxygen white dwarf ends its life accreting matter from a nearby star in binary systems. The only sources of neutrinos are electron capture on free protons and nuclei.  Core collapse supernovae (SNe) are believed to be gravitational collapses of a star heavier than 8 solar masses into a neutron star and 99\% of the binding energy of the star is released in neutrinos. In the early phase ($\sim 10$ ms), called deleptonization or neutronization, about $10^{51}$~ergs are emitted in electron neutrinos from electron capture ($e^{-} + p \rightarrow n + \nu_{e}$) and subsequently ($\sim 10$ s) all neutrino flavors are produced in reactions such as $e^{+} + e^{-} \rightarrow \nu + \bar{\nu}$ in the thermalization phase. 
SN collapse neutrinos can be detected in scintillator detectors such as Baksan \cite{baksan} and LVD \cite{LVD} where the main reaction is inverse-$\beta$ ($\bar{\nu}_{e} + p \rightarrow e^{+} + n$) which gives a prompt signal due to the positron
(visible energy
$E_{\rm vis} = E_{\bar{\nu}_{e}} - Q + m_{e} = E_{\bar{\nu}_{e}} - 0.789$ MeV) 
followed by the signal from the neutron capture $n + p \rightarrow d + \gamma$ ($E_{\gamma} = 2.23$ MeV). LVD is also sensitive to interactions on carbon and iron nuclei \cite{LVD}. It has been suggested to add NaCl to the scintillator that would enhance the event rate especially in scenarios where the average electron neutrino is 30-40 MeV such as the rotating collapsar model \cite{olga}. 
Cherenkov detectors such as Super-Kamiokande \cite{SKcollapse} and IceCube \cite{IceCubecollapse} would also detect the light produced by positrons from inverse-$\beta$ decay. While Super-Kamiokande would detect about $10^4$ events from a SN at 10 kpc equivalent to SN1987A including information on events energy, IceCube, with 4800 optical modules at much larger distances, would detect the average glow of the ice produced by positrons induced by the burst of neutrinos. The advantage of ice with respect to sea water, is the low background rate ($\sim 280$ Hz in IceCube) that mainly depends on the radioactivity from the optical module and phototubes (PMTs) materials. In ANTARES rates are of the order of 100 kHz due to bioluminescence and $^{40}K$ $\beta$-decays. 
On the other hand, in the investigated sites for the future proposed cubic-km scale detector in the Mediterranean, the background rate is about a factor of 2 or more smaller and SN collapse searches 
are possible \cite{sn_km3net}.
In IceCube optical modules (OMs) are read out in bins of 1.6384 ms and the average rate increase per Digital OM (DOM) is about 13 Hz for a SN at 7.5 kpc of distance. Summing over all DOMs this implies a rate of about $10^6$ Hz and hence an evidence of 34$\sigma$. A SN in the Large Magellanic Cloud would still produce
a 5$\sigma$ significant detection.
The mentioned detectors monitoring the Galaxy for SN collapses  send their alerts to SNEWS \cite{snews}.

Current upper limits on the number of supernova collapse in our Galaxy are 
getting closer to the predicted rates of about 2 per century \cite{collapse} (see Tab~\ref{tab1} for a summary of presented results).
What would we learn if a supernova collapse happened in our Galaxy? Surely we would learn about astronomical properties of these events. Moreover we could also extract information on neutrino properties \cite{snosci}. Neutrinos pass through the mantle and envelope of the progenitor star and encounter a vast range of matter densities, implying two MSW resonances: the H-resonance, that corresponds to the atmospheric mass difference, and the L-resonance, that corresponds to the solar one. The H-resonance is particularly interesting because it occurs in the neutrino sector for the normal mass hierarchy, and in the anti-neutrino one for the inverted hierarchy. It is adiabatic for sin$^2 \theta_{13} \gtrsim 10^{-3}$ and non-adiabatic for sin$^{2} \theta_{13} \lesssim 10^{-5}$. IceCube has some sensitivity to $\theta_{13}$ for galactic SN \cite{IceCubecollapse} during the deleptonization burst that is roughly independent on properties of the progenitor star. It is also shown in Ref.~\cite{raffelt} that the signal onset can be reconstructed within $\pm (6-7)$ ms in IceCube at 1$\sigma$ c.l. and this measurement, that can be performed also by other experiments, can also be proved by a coincident gravitational wave detection.  A SN can be 
located using triangulation between detectors as well as the electron recoil in Super-Kamiokande or a future Mton detector as proposed in Europe and in USA even if it cannot be located by astronomical means because of obscuration~\cite{beacom}.

A Mton detector would also play a major role for the search for nucleon decay. At the conference the most interesting current limits were presented by Super-Kamiokande \cite{nucleondecay}. One of the general features of Grand Unification Theories is that they allow lepton and baryon number violations and predict the instability of nucleons. The favored decay mode in GUTs based on SU(5) symmetry is $p \rightarrow e^+ + \pi^0$, while for models that incorporate supersymmetry it is $p \rightarrow \bar{\nu} + K^+$. While SU(5) has been already ruled out, SO(10) predictions for the lifetimes are around $10^{35}$ yrs. With an exposure of 141 kton year, Super-Kamiokande has set a 90\% c.l. limit  on  $p \rightarrow \bar{\nu} + K^+$ lifetime of $2.8 \times 10^{33}$ yrs, where the prediction for SO(10) is $10^{32 \div 34}$ yrs.
On the other hand, the lower limit for $p \rightarrow e^+ + \pi^0$ is $8.2 \times 10^{33}$ yrs, 5 times longer than the previous best limit. Candidate events are consistent with the expected backgrounds from atmospheric neutrinos.
Limits for other modes range between $3.6 \times 10^{31}$ to $6.6 \times 10^{33}$ yrs.
\begin{table}
\caption{\label{tab1}
Summary of current limits on the SN collapse rate}
   \begin{tabular}{cccc}
Detector & Livetime & Active mass & Limit (SN/century) \\
                &      (yr)    &         (ton)   & (90\%cl) \\           \hline
Baksan \protect\cite{baksan}& 24.7  & 130  &  9.3  ($< 10$ kpc) \\
LVD \protect\cite{LVD}& 14.5  & 950  & 15 ($< 20$ kpc) \\
SK 1+2+3 \protect\cite{SKcollapse}& 7.09  & 50,000 &  29 ($<100$ kpc) \\
\end{tabular}
\end{table}

\section{Dark matter and and anti-matter searches}
\label{sec4}

The astronomical evidence for the existence of dark matter has been improving for over 60 years and it is now well established. DM accounts for about 23\% of the universe energy density. We observe anomalies in astrophysical systems ranging from galactic to cosmological scales that can be accounted for if a large amount of not luminous matter is assumed  \cite{bertone}. Gravitational lensing and X-ray observation have disfavored most of the modified-gravity  alternatives to dark matter. Supernova data imply that the expansion rate of the universe is accelerating. If this is not an indication of a breakdown of General Relativity, this implies the existence of  a ``dark energy'' that dominates the energy  density of the universe, fills space and exerts repulsive gravity. Cosmic microwave background measurements have confirmed the deduction from Big Bang nucleosynthesis that the dark matter must be non-baryonic through the detection of the higher acoustic peaks \cite{wmap}. The Lambda Cold Dark Matter model is remarkably successful and the key cosmological parameters are measured with multiple techniques to better than 10\%.
The nature of the non-baryonic particles it is still unknown. Particle theory suggests as possible solution to the DM problem weakly interacting massive particles (WIMPs). If a new particle with weak-scale interactions exists, then its annihilation cross section would be of the right dimension to account for the dark matter in the universe.
Possible DM candidates are axions, that would provide a solution to the CP violation problem, or  supersymmetric particles such as the well motivated neutralinos, or  as well as Kaluza-Klein states which appear in models of universal extra-dimensions.

Dark matter searches are divided in 2 main categories: direct searches where the DM particle scatters elastically in a detector and the nuclear recoil is detected and indirect searches that detect the secondary radiation produced in DM annihilation. 
At this conference indirect detection using gammas ad neutrinos was discussed. Dark matter particles such as neutralinos can annihilate and produce fermion-antifermion pairs, gauge boson pairs and final states containing Higgs bosons. The subsequent hadronization results in a gamma-ray power-law spectrum with a sharp cut-off at the neutralino mass expected to be between about 50 GeV and a few TeV. Direct annihilation in gamma-rays ($\chi \chi \rightarrow \gamma \gamma$  or $Z^0 \gamma$) produce emission lines that would represent a very clear signature, but these processes are loop-suppressed since tree-level Feynman diagrams are forbidden.
Fermi-LAT did not detect any $\gamma$-flux signal at 5$\sigma$ level in 3 months of data and set upper limits at the level of $5-10 \times 10^{-10}$ cm$^{-2}$ s$^{-1}$ sr$^{-1}$ for galactic latitudes  $|b| > 10^{\circ}$ between 50-300 GeV \cite{glast_line}. The mission was successfully launched on June 11, 2008. The Large Area Telescope (LAT) is an electron-positron pair production telescope made of a silicon tracker, a calorimeter and an anti-coincidence system to reject charged particle background, sensitive between 20 MeV and about 300 GeV.  

Indirect detection of DM is mainly affected by uncertainties on astrophysical parameters. The expected gamma-ray flux from DM annihilation is made of 2 terms. The astrophysical factor is proportional to the square of the DM density profile, that represents an important uncertainty in flux estimations of up to 2 orders of magnitude, convoluted with the resolution of the telescope. Many of the works presented at the conference use the Navarro-Frenk-White profile \cite{nfw}. Nonetheless, it should be noted that gamma fluxes are particularly sensitive to any DM enhancements due to the presence of substructures in the halo. The other term is a particle physics factor that depends on the inverse of the squared neutralino mass, the annihilation cross section times the velocity of the DM, and the gamma-ray annihilation energy spectrum. The allowed parameter space for the mass and the annihilation cross section spans many orders of magnitude resulting in estimations which can differ up to 6 and more orders of magnitude. Various regions of interest have been looked up by Imaging Atmospheric Cherenkov Telescopes (IACTs). Promising targets with high DM density relatively close to the Earth ($\gtrsim 100$ kpc) are dwarf Spheroidal (dSph) satellite galaxies of the Milky Way whose dynamics are dominated by DM halos of the order of $10^5-10^9 M_\odot$, very high mass-to-light ratios (up to $\sim 10^3 M_{\odot}/L_{\odot}$) and with no astrophysical gamma-ray source in the vicinity. Other sources of interest are clusters of Galaxies, that are the most massive gravitationally bound systems in the universe, with radii of the order of the Mpc and total masses of around $10^{14}-10^{15} M_{\odot}$, and intermediate mass black holes.
The 17-m telescope MAGIC-I, located at 2200 m a.s.l. in the Canary Island of La Palma, set limits for the dSph Draco and William 1 \cite{MAGIC_dm}. These limits are still far to constrain the mSUGRA parameter space. When MAGIC will operate in the stereo mode, DM searches will profit of the lower energy threshold and better background discrimination at low energies. VERITAS 
set limits on dSph galaxies and globular clusters such as M5, M32 and M33 \cite{VERITAS_dm}. William 1 limits are about a factor of 2 better than MAGIC-I ones in the region of $2.2 \times 10^{-12}$ cm$^{-2}$ s$^{-1}$ (95\% c.l.) but still far from mSUGRA predictions by about 4-5 orders of magnitude. H.E.S.S. \cite{HESS_dm} looked in the direction of  Canis Major, an overdensity that could be a dwarf galaxy or a part of the warped Galactic disk. Upper limits were set for Kaluza-Klein models and phenomenological Minimal Supersymmetric extensions of the Standard Model. H.E.S.S. past observations disfavored DM emission from the Galactic Centre and revealed instead a VHE source (HESS J1745-290)\cite{Aharonian06}.

Fermi-LAT performance for DM detection are considerably better than for IACTs thanks to the lower energy threshold. A preliminary study for the DM emission from the Galactic Centre has been presented: various gamma sources have been detected in the Galactic Centre region, hence a DM analysis should subtract their contribution \cite{fermi_galcentre}. Limits for some dSph galaxies have also been set using 9 months of data and begin to constrain some MSSM models. At 100 GeV of neutralino mass limits on $<\sigma v>$ are between $10^{-25}-10^{-24}$ cm$^3$ s$^{-1}$, many orders of magnitude better than IACT limits, while at 1 TeV they are comparable to IACT limits at around $10^{-23}$  cm$^3$ s$^{-1}$ \cite{glast_dwarf}. 

\begin{figure*}[!t]
   \centerline{\subfloat [Upper limits (90\%c.l.) on the muon fluxes from neutralino annihilation in the Sun as a function of neutralino mass for soft ($b\bar{b}$) and hard ($W^+W^-$) channels. The lighter hatched region are SUGRA models compatible with direct detection limits on the spin independent cross section $\sigma^{SI}$ from CDMS \protect\cite{cdms} and XENON10 \protect\cite{xenon}. The darker hatched region is the same but not disfavored if direct detection limits were 100 better.
 Experimental limits, that include a correction for the threshold of the detectors for the common assumption of $E_{\nu,thr} = 1 GeV$, are shown for Super-Kamiokande \protect\cite{dm_sk}, MACRO ref\cite{macro_dm}, AMANDA-II \protect\cite{dm_amanda}, Baikal and IceCube 22 strings for  the hard and soft channels. Also ANTARES sensitivity in 5 yrs is shown.]{\includegraphics[height=2.5in,width=2.9in]{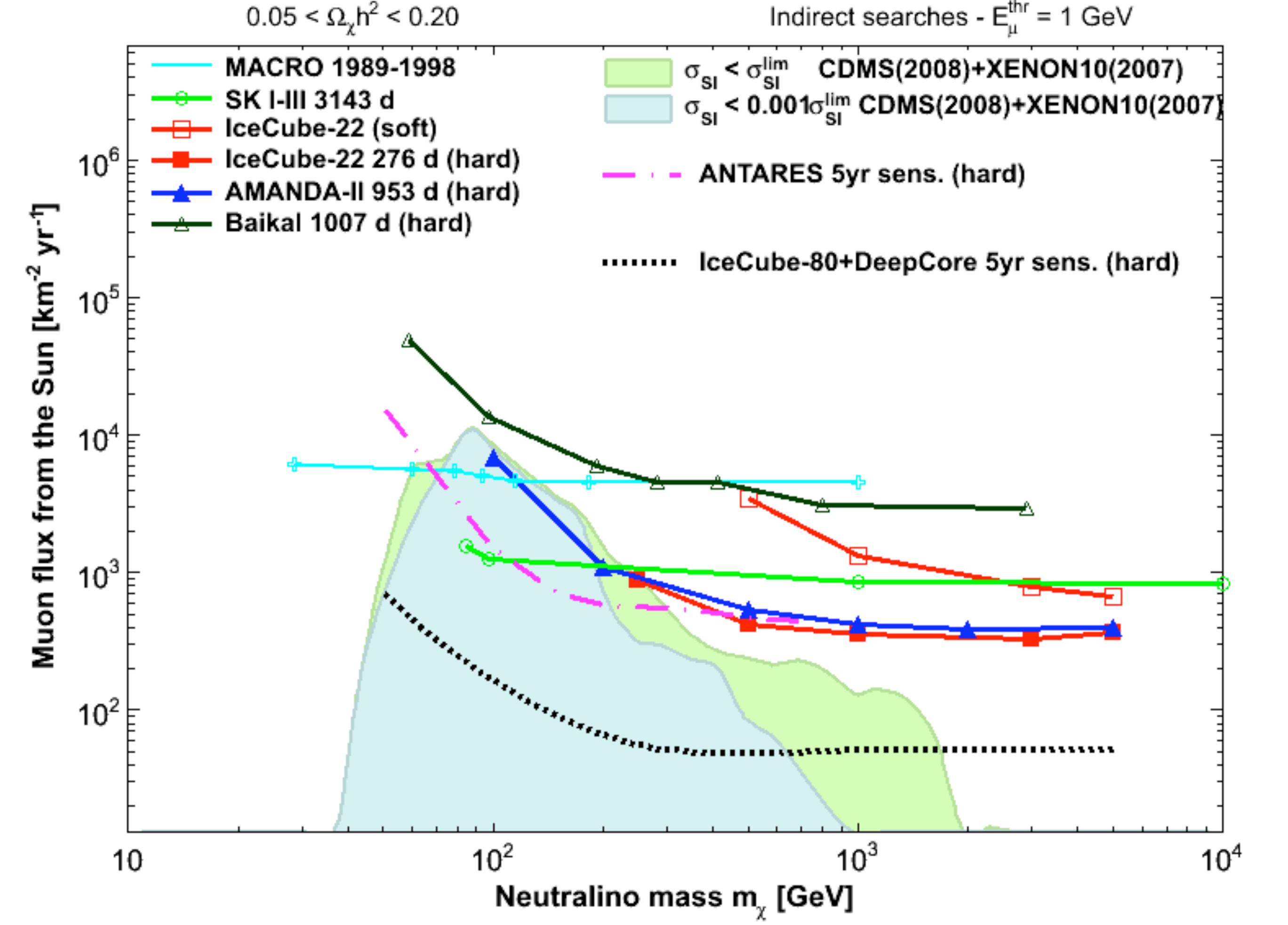} \label{left_fig3}}
              \hfil
             \subfloat[90\% c.l. upper limits for AMANDA-II and 22 strings of IceCube on the spin dependent cross section for soft and hard channels. Hatched regions follow the same conventions as in Fig.\protect\ref{left_fig3}. Also shown are direct search limits frm CDMS \protect\cite{cdms}, COUPP \protect\cite{coupp}, KIMS \protect\cite{KIMS} and Super-Kamikande \protect\cite{sk_dm1}.]{\includegraphics[height=2.5in,width=2.9in]{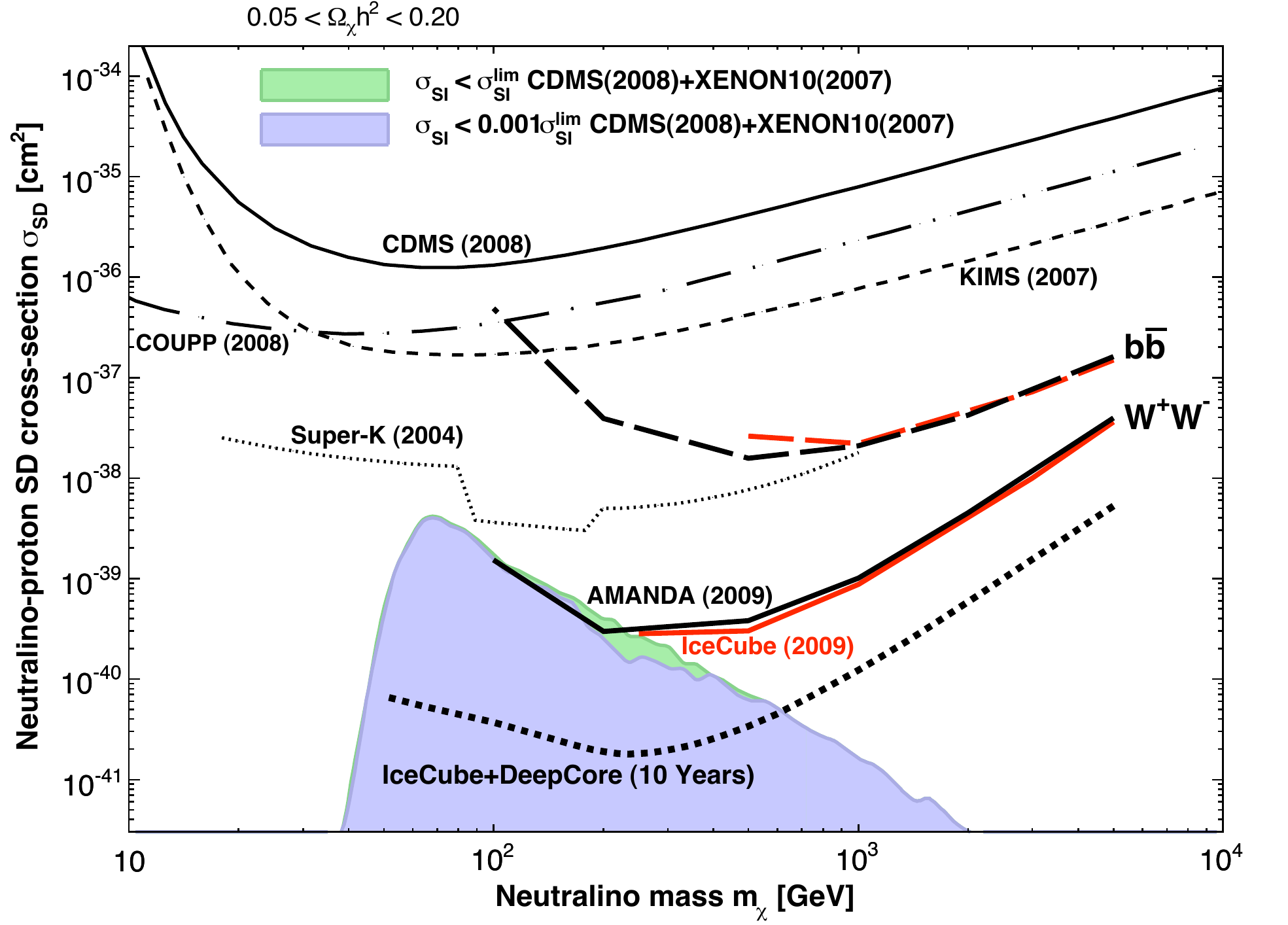}\label{right_fig3}}
            }
   \caption{}
   \label{fig3}
 \end{figure*}

While gamma-ray experiments set limits on the annihilation cross section times the DM velocity or the gamma-flux, neutrino telescopes measure the muon induced flux by neutrinos and constrain this flux as a function of the DM particle mass. WIMPs may become gravitationally trapped in celestial bodies like the Sun or the Earth or the Galactic Centre and would annihilate producing neutrinos between other secondaries. The capture rate depends on the elastic scattering cross section of the WIMP on the nucleons in the celestial body, e.g. the Sun is dominated by H atoms and so WIMPs predominantly undergo Spin Dependent interactions (SD). This is the reason why indirect searches with neutrinos perform much better than direct searches if WIMP SD interactions are dominant. This indicate that these searches cover complementary parameter spaces of various models. The capture rate also depends on the local density of the dark matter ($\rho_{\rm loc} \sim 0.3$ GeV/cm$^3$), on the local rms velocity of the DM ($\sim 270$ km/s) and on the inverse squared mass of the WIMP. If the capture rate $C^{\odot}$ and the annihilation cross sections are sufficiently high, equilibrium may be reached between these 2 processes and for N WIMPs in the Sun the rate of change is $\dot{N} = C^{\odot} - A^{\odot} N^2$,  where $A^{\odot} = \frac{<\sigma v>}{V_{\rm eff}}$ is the annihilation cross section times the velocity of the WIMP.  Gamma-experiments constrain $<\sigma v>$ and $V_{\rm eff}$ is the effective volume of the core of the Sun that depends on the mass of the gravitationally trapped WIMP.
This flux of resulting neutrinos is proportional to the annihilation rate $\Gamma = 1/2 A^{\odot} N^2$ and it is maximal when equilibrium between annihilation and capture is reached \cite{bertone}.
Neutrino telescopes at these conference looked for an excess of events with respect to the atmospheric neutrino background mainly from the direction of the Sun, and also from the Earth core and the Galactic Centre. They set limits on the muon induced flux of neutrinos and showed how these limits constrain the interaction cross section so that a comparison with direct searches is possible. A collection of various results including those presented at this conference is shown in Fig.~\ref{fig3}. The conversion from muon flux to cross section limits is not model independent and it is done using DarkSUSY \cite{darksusy}. Equilibrium is assumed between capture and annihilation rates in the Sun, so that the annihilation rate is proportional to the spin-dependent and independent cross sections. A limit on $\sigma^{SD}$ is found setting to zero the spin-independent cross section. This procedure is indispensable to show complementarity between searches and combine results from different techniques. This step is still missing between gamma indirect searches and neutrino ones or direct searches.
This method was applied by IceCube \cite{dm_ic22}, and previously by Super-Kamiokande \cite{sk_dm1}. IceCube performed a search using the 22 string configuration. Further improvements will be possible with the full detector and Deep Core, a central, tighter 6 string array instrumented with high quantum efficiency PMTs (see Sec.~\ref{sec5}). Deep Core will enhance IceCube sensitivity in the region between 20-300 GeV, also important for DM searches. IceCube also presented a search for the lightest and stable Kaluza-Klein state \cite{kk_ic22} from the Sun. Limits for the final sample of AMANDA, that has been decommissioned this year, are comparable with the 22 strings configuration but go down to lower energies \cite{dm_amanda}. Other limits of interest were presented by Baikal \cite{dm_baikal} and Super-Kamiokande \cite{dm_sk}. ANTARES \cite{dm_antares} derived first limits for 5 lines data, that are yet not competitive with other experiments, while expected ones are particularly interesting in the low energy region (see Fig.~\ref{fig3}). 

Antimatter searches were mostly covered in other sessions \cite{sinnis}. Best upper limits on the $\bar{He}/He$ ratio were presented by the BESS spectrometer that had two polar flights: Polar I  of 8.5 d in 2004 and Polar II of 24.5 d in 2007-8 \cite{bess}. They analyzed the data of Polar I and 1/3 of the data of Polar II and the resulting upper limit (95\% c.l.) is $1.5 \times 10^{-7}$ in the 1-14 GV rigidity range. With the full statistics the limit will be improved by almost an order of magnitude.

AMS-02 will have an impact on anti-matter searches, given the wider energy range (up to 1 TeV)  that can now be covered by PAMELA \cite{PAMELA}. AMS-02 is expected to be launched in Sep. 2010 and to spend 3 yrs on the International Space Station \cite{ams}. It will measure the electron and positron spectra with much better statistics than PAMELA (that has an acceptance of about a factor of 200 smaller) and AMS-01. If no antimatter will be observed by AMS-02 we will probably be able
to exclude the presence of antimatter up to about 1000~Mpc.

\section{Neutrino Telescope results on searches for astrophysical neutrino sources and cosmogenic neutrinos}
\label{sec5}

It took about 35 yrs since Markov conceived the Neutrino Telescope (NT) detection principle \cite{Markov:1960} to the operation of the first complete detector.  NTs are challenging instruments to build, while they use the well established technique of photomultipliers.
The difficulties for construction arise from: the necessity of instrumenting extremely large regions due to the small neutrino cross section and to the steep decrease of incoming neutrino fluxes of atmospheric and astrophysical origin; the darkness and transparency of the medium requirements to detect the faint Cherenkov light produced by ultra-relativistic charged particles; the need for filtering neutrino events out of the more abundant atmospheric muons. Detectors are located under 1-4 km of water or ice where the downward-going background of atmospheric muons is  about $5 \div 6$ orders of magnitudes larger than the atmospheric neutrino flux for $E_{\nu} \gtrsim 100$~GeV  coming from all directions (see Fig.~\ref{fig2}).

NTs are tridimensional matrices of photodetectors made of strings of optical modules that  house large photocathode PMTs and protect them from the water column pressure when they are installed in the sea or in lakes and from the pressure during ice refreezing for antarctic detectors. Neutrinos interact with the medium nuclei in and around the detector through charged and neutral current interactions and charged relativistic secondaries produce light. Various event topologies can be identified: muons, cascades and composite events made of tracks and cascades. 
In the case of cascades, that are point-like events given the scale of these sparse detectors, light radiates almost isotropically. Analyses using cascade topology are at a less advanced stage in IceCube and ANTARES, while Baikal presented very interesting results on this topology \cite{baikal}. In fact, because of the limited dimension of the detector and  the shallow depth of 1.1 km, reconstruction of muons is hampered by the limited number of degrees of freedom and the larger atmospheric muon background. The latest configuration of the Baikal experiment (NT-200) was put into operation in April 1998. It consists of an umbrella-like structure of 8 (72~m long) strings, with 24 pairs of up-looking and down-looking PMTs (containing 37-cm photocatodes developed for the project, called QUASAR-370). Three external strings at 100~m from the center of NT200, each with 12 pairs
of OMs, have been added in April 2005 to increase the cascade sensitivity at very high energies (NT200+). 

\begin{figure*}[!t]
   \centerline{\subfloat[Topologies induced by tau neutrinos: inverted-lollipop and lollipop with a shower followed by a track or a track followed by a shower (hardly distinguishable from a $\nu_{\mu}$ CC interaction and a muon, even if $\mu$ and $\tau$ energy losses are different); double-bang (described in the text). ]{\includegraphics[width=2.9in, height=1.7in]{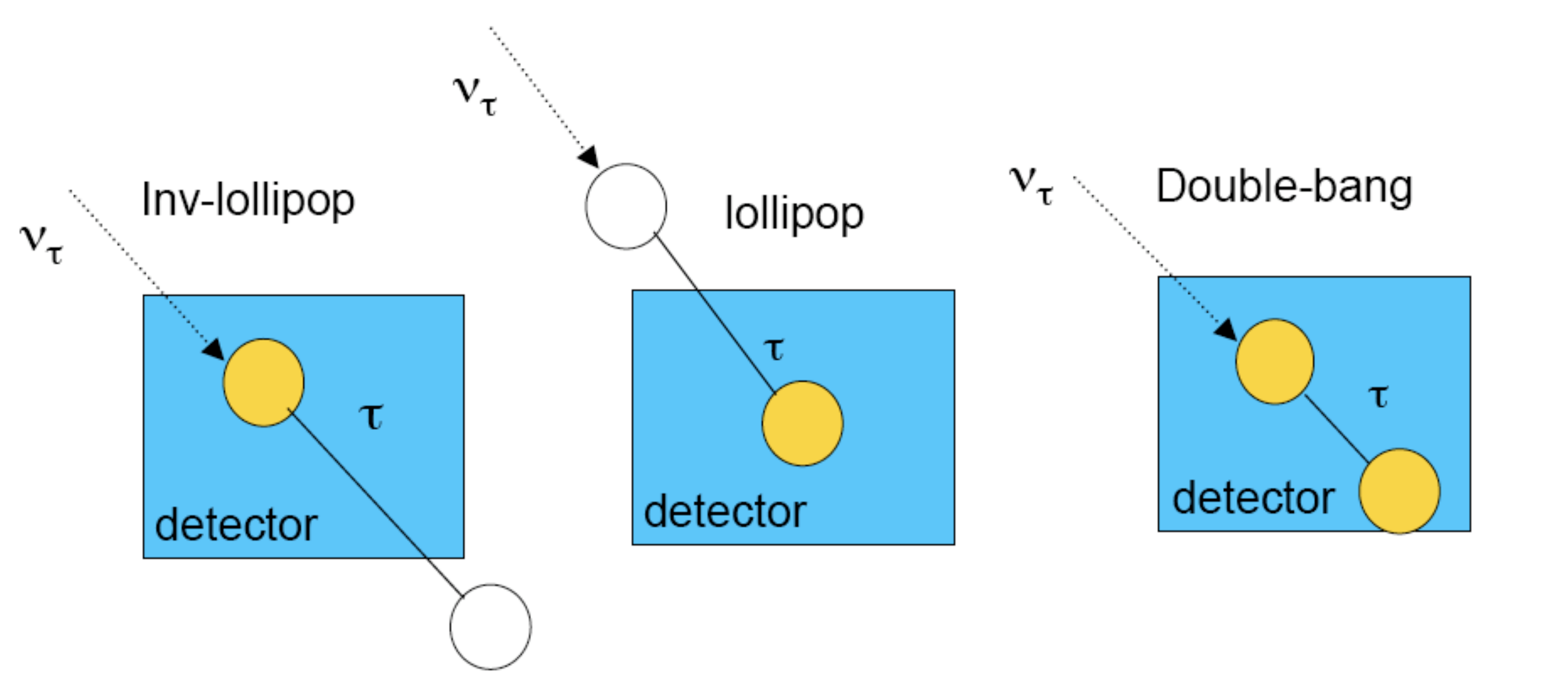} \label{left_fig4}}
              \hfil
              \subfloat[Differential number of events as a function of energy (response curve) for
              40 strings of IceCube for the point-source analysis in the Northern hemisphere (up-going events) and in the Southern one (down-going events) for atmospheric neutrinos, $E^{-2}$ and $E^{-1.5}$ spectra. NTs effective area increases steeply with energy due to the cross section and muon range increase. Hence, the response curve of the detector depends on the spectrum of neutrinos and on analysis cuts.~\protect\cite{Dumm}.]{\includegraphics[width=2.9in]{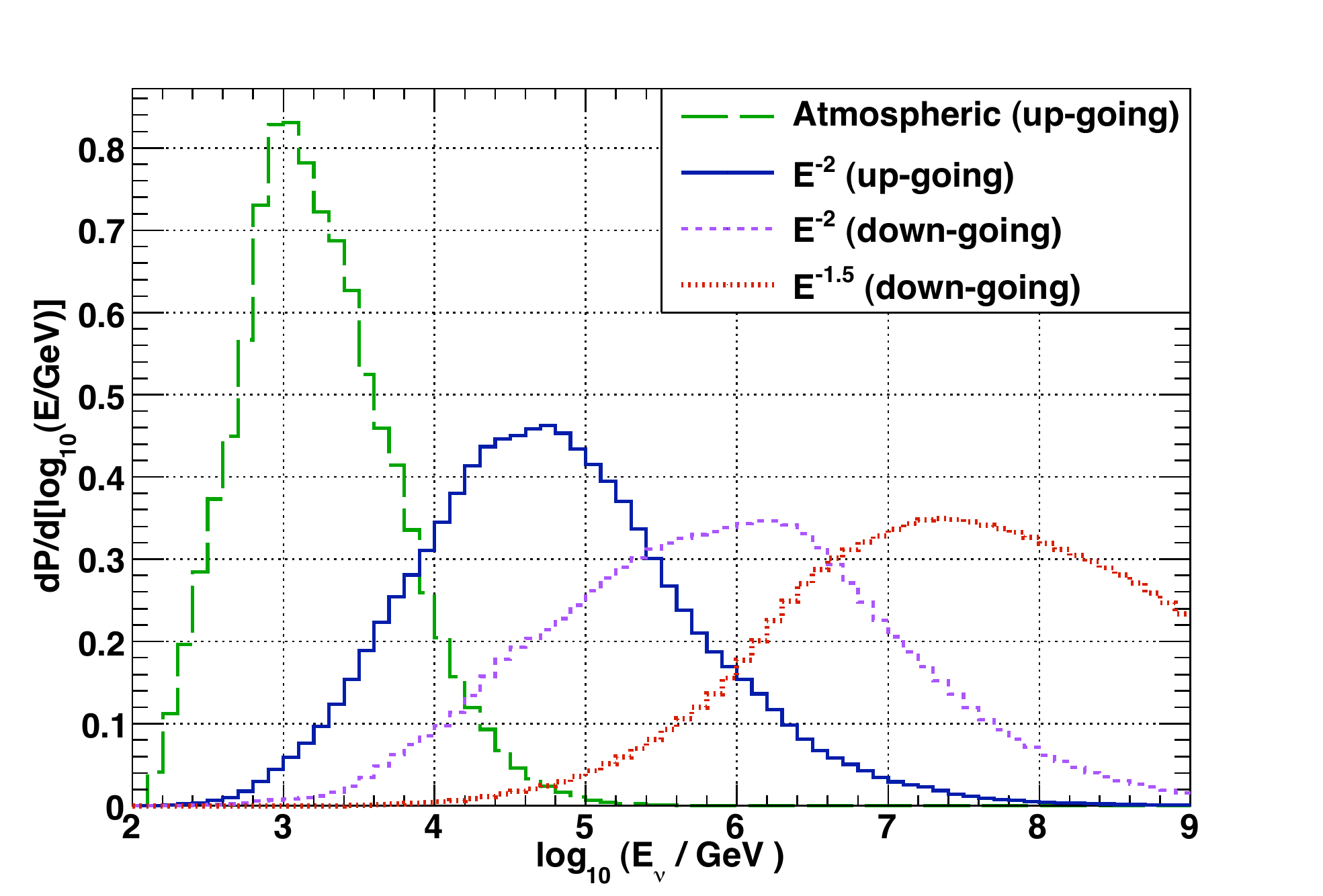} \label{right_fig4}}
             }
   \caption{}
   \label{fig4}
 \end{figure*}

In IceCube, reconstruction of neutrino induced cascades is challenged by scattering of light in the ice. The slight anisotropy in the direction of high energy primary neutrinos should allow a resolution of the order of $20^{\circ}$ while the energy resolution is $\Delta($log$_{10} E) \sim 0.3$ between 20 TeV and a few PeV \cite{ic22_joanna}.
Other topologies are shown in Fig.~\ref{left_fig4} and they are induced by
tau neutrinos \cite{ic22_tau}. One of these is considered background free and called ``double-bang'' due to the presence of two separated cascades induced by the hadronic interaction of $\nu_{\tau}$ and the hadronic or electromagnetic shower produced in the tau decay. In order to reconstruct 2 cascades separately in a detector like IceCube the tau track must be at least of the order of 100 m. Hence this topology is limited in energy to the range of a few PeV for the tau track to be long enough, up to about 100 PeV to contain the two cascades in the instrumented volume (the $\tau$ range is 1 km at about 200 PeV).

Muons propagate along lines for long distances and constitute the ``golden sample'' for neutrino astronomy since the achievable angular resolution is at sub-degree level. At energies $\gtrsim 10$ TeV the kinematic angle between $\nu_{\mu}$ and secondary muons is negligible with respect to the intrinsic tracking resolution due to scattering of light in the Cherenkov medium and to the time resolution of PMTs. Hence, muons can be used to point back to neutrino astrophysical sources. 
While IACTs can prove their point spread function on copious sources of gammas such as the Crab Nebula, no astrophysical neutrino source has proved to exist yet. Hence, the `standard candle' for neutrino telescopes can be an `anti-source', such as the Moon shadow. The Moon disk, with a diameter of about $0.5^{\circ}$, blocks primary cosmic rays hence producing a deficit in the muon flux. IceCube, using 3 months of data of the 40 string configuration, begins to find evidence at $5\sigma$ level of this deficit, which is a proof of the absolute pointing capability of the detector \cite{moon}. The Moon at the South Pole is never higher than $30^{\circ}$, so this tests an important region for the expected astrophysical neutrino signal.
In fact, the astrophysical signal (with characteristic spectrum of $E^{-2}$ from acceleration mechanisms at sources) begins to emerge from the more steeply falling atmospheric neutrino background above about 1-10 TeV. At PeV energies the Earth shadowing effect on neutrinos is no more negligible, hence at very high energy most of the neutrinos can only be seen in the horizontal region. IceCube is able to 
test the angular resolution for less inclined events also using events in coincidence with the extensive air shower (EAS) IceTop. ANTARES is instead evaluating the possibility to measure coincident events using a scintillator array on a boat.

The IceCube observatory consists of a deep detector with instrumented strings between 1.5 and 2.5 km below the surface and IceTop made of two surface tanks at a distance of about 10 m corresponding to each deep string \cite{karle}. In 2 austral summers at the South Pole, construction will be completed for a total of 86 strings, 160 tanks, 4800 10" inch PMTs and 360 high quantum
efficiency PMTs for the Deep Core. In the final configuration the angular resolution at energies between
10-100 TeV will be such that about 50\% of the reconstructed muons from the direction of a neutrino point source will be inside $0.5^{\circ}$. A better angular resolution at the level of $0.2^{\circ}$ should be achievable in sea water as indicated by simulation studies with ANTARES \cite{coyle}.
The construction of the undersea neutrino observatory ANTARES, located
at about 2.5~km below sea surface off-shore Toulon in South France,
took several years. Since 1996, the Collaboration has deployed many strings to monitor the environmental parameters of water, the permanent electro-optical cable of about 40~km, transmitting data and power between the shore station in La Seyne sur Mer, and the junction
box off shore that is operating since November 2002.  In 2006, the first 2 lines of the
detector were deployed and 8 additional lines were disposed in 2007.
On May 2008 the complete NT, made out of a total of 12 lines,
was put into operation. Lines were connected during 5 submarine
operations, one conducted by a submarine and all others by an unmanned
Remote Operated Vehicle. The ANTARES observatory comprises 12
mooring lines, a line specifically dedicated to
marine environmental monitoring, a seismometer, and a biocamera for
bioluminescence studies.  The lines are anchored to sea bed at 2475~m
depth and held vertical by buoys. Buoys are freely floating so each
line moves under the effect of the sea current, with movements of a
few meters for typical values of 5~cm/s. An acoustic positioning
system, made of transponders and receivers, gives a real time
measurement of the position of the OMs with a precision better than
10~cm, typically every 2 minutes and tiltmeters and compasses provide
their orientation \cite{acoustic}.  Seventy-five OMs along each lines between about
2400 and 2000~m are grouped in triples on storeys. The 3 PMTs
(Hamamtsu 10") look downward, at $45^{\circ}$ from the
vertical, to prevent transparency loss due to sedimentation. Storeys also include titanium containers housing the front-end electronics.  Each of the OMs contains a pulsed LED for
calibration of the relative variations of PMT transit time and a
system of LED and laser Optical Beacons allows the relative time
calibration of different OMs.  An internal clock system, which is
synchronized by GPS to the Universal Time with a precision of $\sim
100$~ns, distributes the 20 MHz clock signal from the shore. Time
calibrations allow a precision at the level of 0.5~ns ensuring the
capability of achieving an angular resolution at the level of
$0.2^{\circ}$, for muons above 10 TeV~\cite{calibrations}. All data
above a threshold of about 1/3 of a photoelectron pulse is sent to shore for
further online filtering. This requires coincidences between PMTs on the
same storey and coincidences between hits  compatible with
light propagation in water~\cite{daq}.

\subsection{Point-like source searches}

\begin{figure*}[!t]
   \centerline{\subfloat[Muon event skymap ($z$ axis is relative intensity) for
  six months of data taken with the 40-string IceCube array, from July
  2008 through December 2008. Of the 17777 black dots on the skymap,
  6797 are up-going neutrino candidates (zenith $> 90^\circ$) from the
  northern hemisphere. The color shading indicates the significance of
  the data and the curved black line is the galactic plane~\protect\cite{Dumm}.]{\includegraphics[width=3.in,height=2.2in]{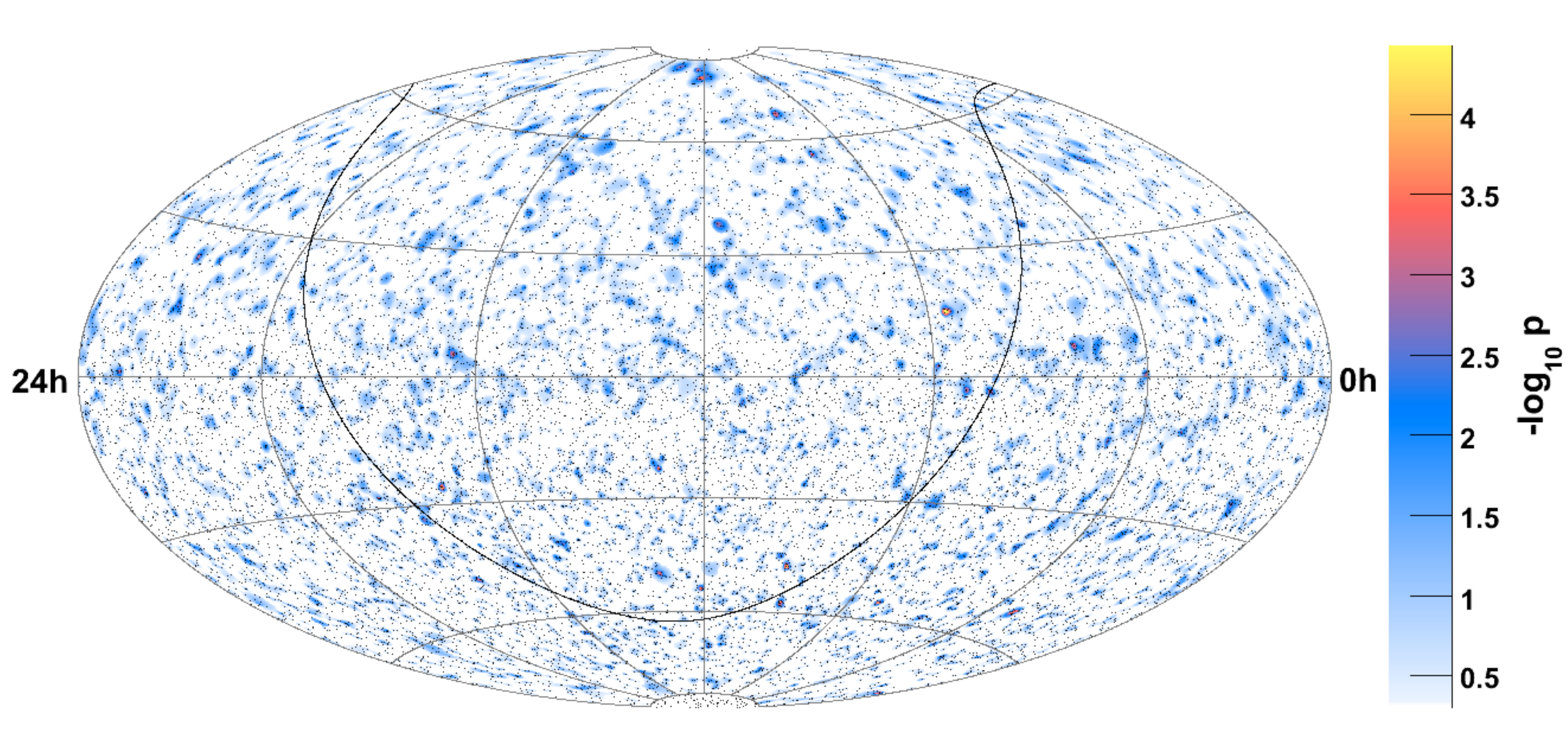} \label{left_fig5}}
              \hfil
              \subfloat[Number of signal events from a point source to fit the spectral index of a source at declination of $6^{\circ}$. Shaded regions are $1\sigma$ errors and the `true' spectral index is indicated on the right.]{\includegraphics[width=3.in,height=2.4in]{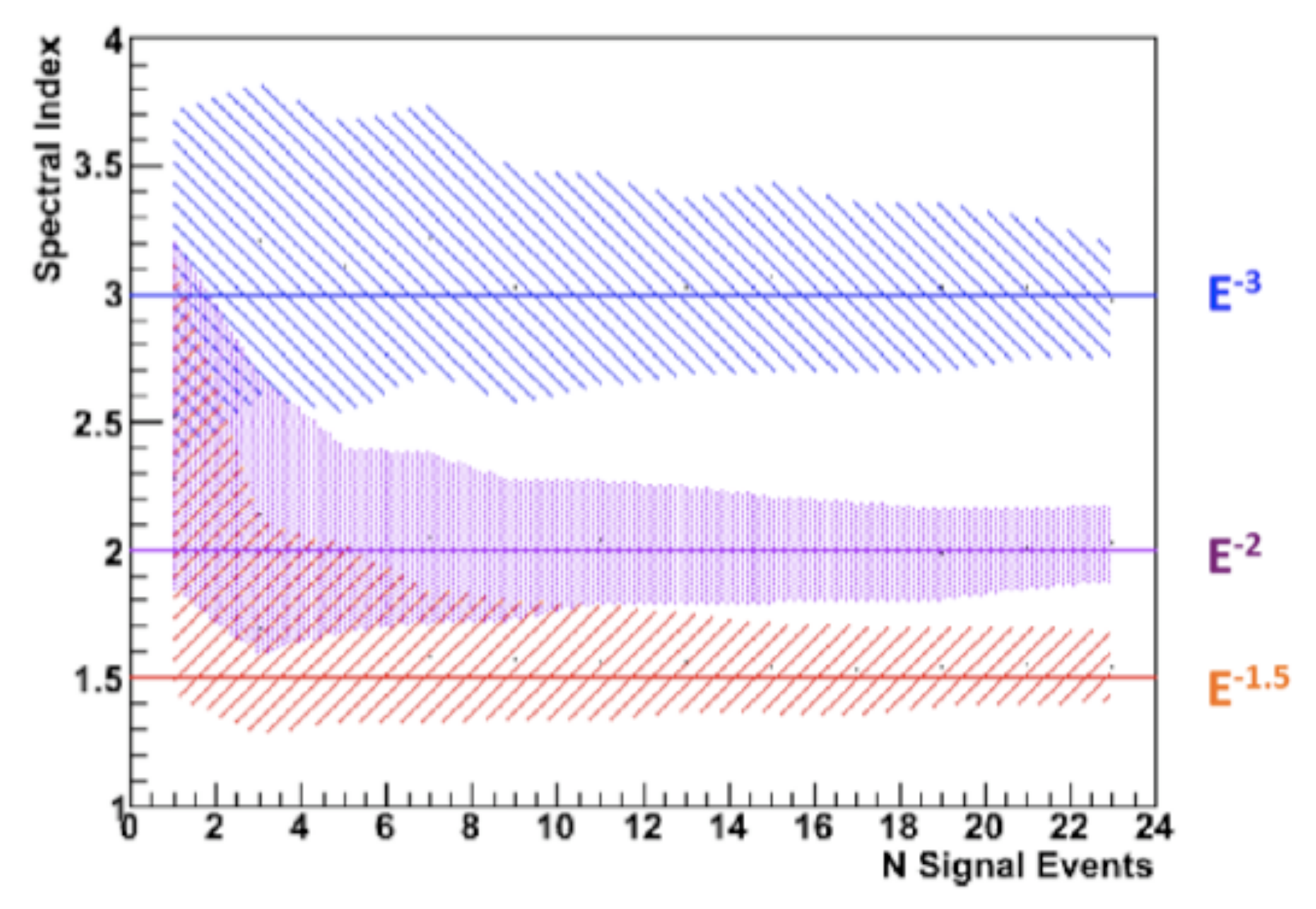} \label{right_fig5}}
             }
   \caption{}
   \label{fig5}
 \end{figure*}

Point-like source searches have various advantages compared to diffuse ones. First of all, the directionality of the signal helps to discriminate the background of atmospheric neutrinos. Many `equivalent experiments' with only background can be reproduced by scrambling the right ascension of data. This has the advantage that data can be used for estimating the background rather then simulation, so that resulting significances are not affected by imperfect understanding of the detector and detector medium, of signal and background. Likelihood methods \cite{method} are used to discriminate between the signal + background hypothesis and the background only one. Signal is characterized with respect to background by the source directional feature and by the harder spectrum expected from astrophysical sources compared to atmospheric neutrinos. For the case of time dependent sources, such as periodical binaries as micro-quasars or flares from blazars, also periodicity assumptions or light curves from gamma, X-ray and optical telescopes can be used as parton distribution functions \cite{mike}. 
IceCube recently published results for the 22 string configuration with 5114 events from the Northern 
hemisphere collected in 276 days and a hot spot was found with a post-trial probability of 1.34\% to be due to background fluctuations \cite{ps_ic22}. The significance was driven to such values by 3 high energy events of a few hundreds of TeV. IceCube field of view (FoV) has been extended to a part of the Southern Hemisphere using energy cuts to reduce the background of atmospheric muons. Hence, in this part of the sky IceCube is sensitive to PeV sources \cite{lauer}.
At the conference, new results from 6 months of data of the 40 string configuration, that encompass in sensitivity all previous samples, have been presented \cite{Dumm}. The statistical fluctuation observed with 22 strings was not confirmed with the larger statistics sample, indicating that the hot spot was a statistical fluctuation.
No signal has been found and upper limits have been set that are the best limits available for the Northern hemisphere.
This analysis extends the FoV of IceCube to the entire sky with the caveat that IceCube is sensitive to TeV-PeV source in the Northern hemisphere where the on-off technique is done using 
atmospheric neutrinos at a rate of 39/day, and to PeV-EeV sources from the Southern hemisphere 
where the method is applied to 62/day high energy atmospheric muons. The response curves of 40 strings of IceCube in the upgoing and downgoing regions and for different spectra are shown in Fig.~\ref{right_fig4}. The sky-map of all events used for point-source searches is shown in Fig.~\ref{left_fig5}. It is also interesting to understand how well the likelihood method would reconstruct the spectral index of a source of different intensities. This is shown in Fig.~\ref{right_fig5}. The relatively good ability of NTs to reconstruct spectra even for a limited energy resolution of $\Delta ($log$_{10}E \sim 0.3$) is due to the wide energy range over which they operate (between about 100 GeV to EeV energies): detected events provides a long lever arm for spectral reconstruction. 

ANTARES, that uses another unbinned method called Expectation-Maximization method \cite{aguilar}, presented the first results with the 5-line configuration \cite{ps_antares}.
A collection of all results for point source searches is presented in Fig.~\ref{left_fig6}.

Various target of opportunity programs (ToOs) are ongoing between neutrino telescopes, optical and 
gamma telescopes as well as with GW interferometers. 
IceCube has activated a program with the ROTSE-III optical telescopes for monitoring high energy neutrino emission from SNe and gamma-ray bursts \cite{icecube_rotse}. Multiplets of at least two muon neutrinos within $4^{\circ}$ arriving in a 100 s time window are sent as alerts to the network of optical telescopes that
look for transient objects. An alert rate of 25 yr$^{-1}$ is estimated. If no SN is detected by 40 strings of IceCube, the rate of SNe with mildly relativistic jets is $< 3 \times 10^{-6}$ Mpc$^{-3}$ yr$^{-1}$ (90\% c.l.). ANTARES is collaborating for a similar program with TAROT and sending multiplets and high energy events \cite{antares_tarot}. Another program involves IceCube and MAGIC: interesting (at $3\sigma$ level the event rate is 2 yr$^{-1}$) IceCube events clustered in time and from selected blazars will be sent to MAGIC \cite{icecube_magic}. Also an analysis of interesting events for ANTARES and GW interferometers, VIRGO and LIGO \cite{antares_virgo}, is being activated with an alert expected rate of 1 every 600 yrs. No one can yet predict the pay-off of these ToO programs between experiments that have yet seen no signal. Traditionally target of opportunity programs have been activated between experiments that know well their signal and the background.

\subsection{Diffuse-flux searches for extra-galactic sources}

\begin{figure*}[!t]
   \centerline{\subfloat[Summary of results on sensitivities and upper limits (90\% c.l.) for $E^{-2}$ neutrino point source fluxes as a function of declination. In the Southern hemisphere from bottom to top: upper limits (red empty squares) for specific sources and sensitivity (red solid line) for 5 lines of ANTARES \protect\cite{ps_antares}; new results from SK \protect\cite{sk_astro} (notice the up-fluctuation for SK for RXJ 1713.7-3946, very close to ANTARES upper limit); ANTARES sensitivity for 1 yr (dotted red line), KM3NeT sensitivity for 1 yr (large dotted red line) \protect\cite{km3net}. It should be noted that in this region IceCube is sensitive to higher energy events than ANTARES (see Fig.~\protect\ref{right_fig4}).
In the Northern hemisphere from top to bottom: AMANDA-II final sensitivity (pink solid line) and upper limits for specific sources (pink circles) \protect\cite{ps_amanda}. The amount of scattering of these points depends on the statistical fluctuation of the background. IceCube 40 string configuration \protect\cite{Dumm}: the upper blue dashed-dotted line is the $5\sigma$ discovery flux for the entire statistics (345 d), the solid line is the sensitivity for the unblinded statistics at ICRC2009 (140 d), and the blue solid line is the sensitivity for the full statistics. The dashed black line is the full IceCube sensitivity for 1 yr for preliminary cuts defined on the 40 string configuration.]{\includegraphics[width=3.in,height=3.in]{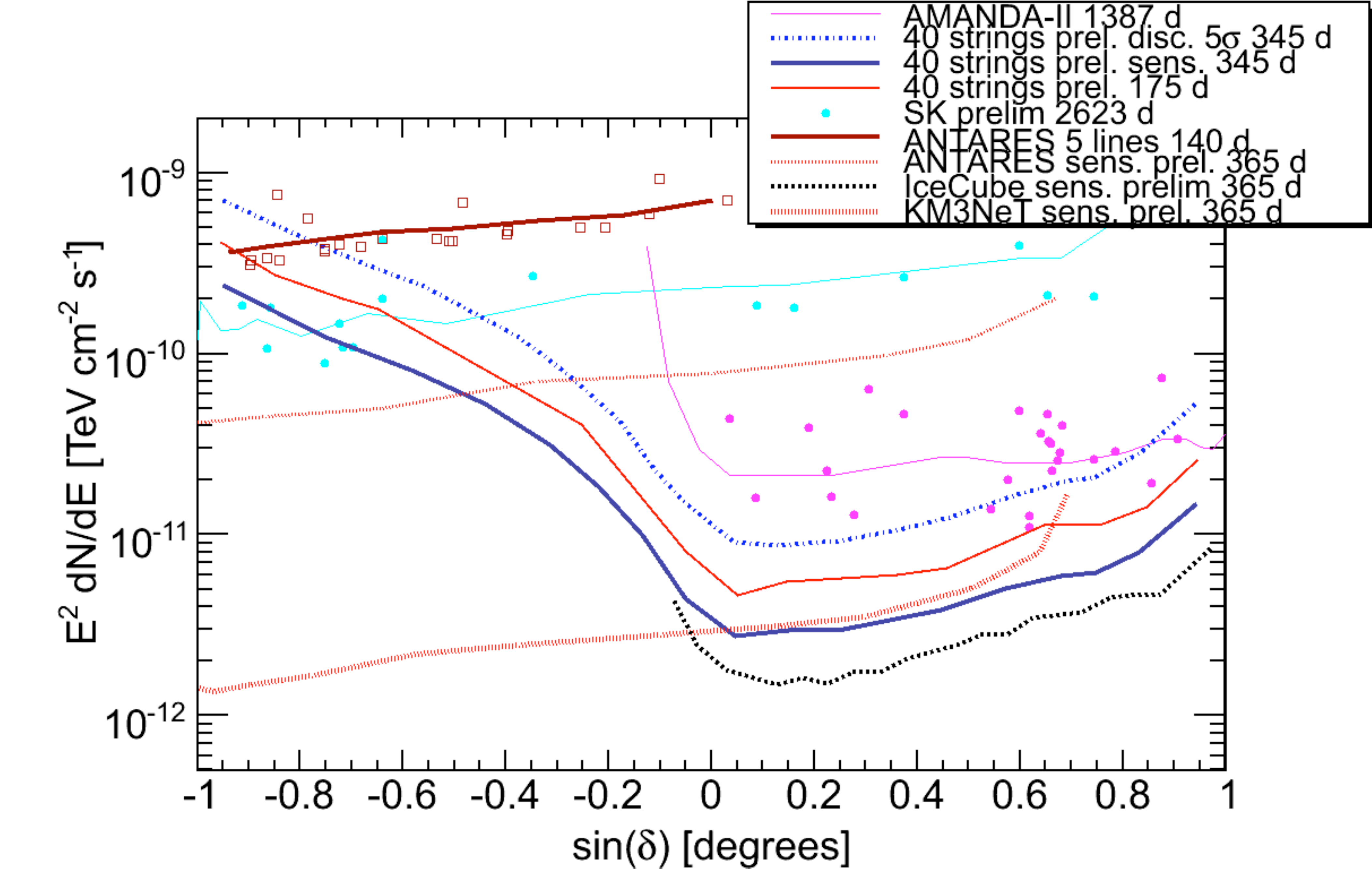} \label{left_fig6}}
              \hfil
              \subfloat[Limits and predictions (90\% c.l.) for diffuse fluxes of neutrinos as a function of energy. Here we assume equipartition of flavors due to oscillations. Though this is a debatable procedure, we multiply by factors the fluxes and limits that are not for all flavors or that do not account for oscillations. Factors are indicated in the legend. On the left (for $E_{\nu} \lesssim 10^8$~GeV): triangles are the atmospheric muon neutrino unfolded spectrum measured by 22 strings of IceCube in 242 d \protect\cite{dima_ic22} compared to a combination of conventional and prompt neutrino calculations (the higher flux is obtained summing the Bartol flux \protect\cite{bartol} and the RQPM model in \protect\cite{naumov} and the lower one summing Honda \etal \protect\cite{honda} and Sarcevic \etal one \protect\cite{sarcevic}). The horizontal solid lines are $90\%$ c.l. limits on $E^{-2}$ neutrino fluxes. From top to bottom: AMANDA-II all-flavor cascade limit for 1001 d \protect\cite{amanda_cascades}, Baikal cascade limit for 1038 d \protect\cite{baikal}, AMANDA-II muon neutrino limit for 804 d \protect\cite{amanda_muon}, all flavor Ultra-High Energy (UHE) limit for 507 d of AMANDA-II \protect\cite{amanda_uhe}. On the right ($E \gtrsim 10^8$~GeV): differential in energy upper limits (90\% c.l.) for 22 strings of IceCube \protect\cite{ic22_ehe}, Pierre Auger $\nu_{\tau}$ \protect\cite{auger}, HiRes \cite{hires}, ANITA \protect\cite{anita}, RICE \cite{rice} and the lunar regolith experiment ATCA \protect\cite{atca}. Models are: Waxman and Bahcall upper limit corrected for oscillations \protect\cite{wb} (black dotted), AGN models (dashed red curves) are from Fig.~20 in \protect\cite{becker} (curves labeled 1 \protect\cite{stecker} and 5 \protect\cite{mucke}), GRB models are from Ref.~\cite{razzaque} for prompt emission and precursor and cosmogenic predictions are from \protect\cite{engel} and the pure proton model in \protect\cite{allard}.]{\includegraphics[width=3.in,height=3.in]{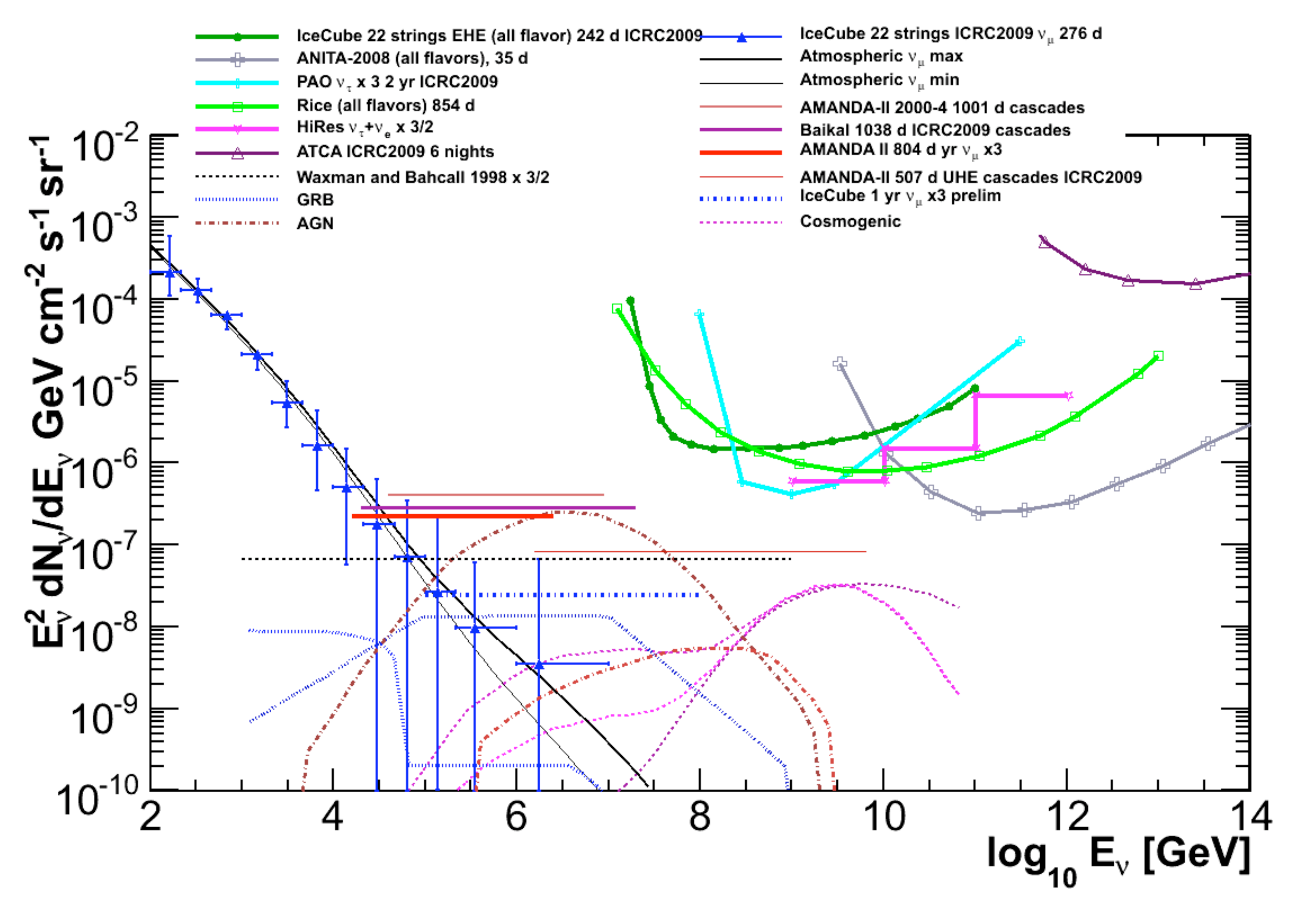} \label{right_fig6}}
             }
   \caption{}
   \label{fig6}
 \end{figure*}

Diffuse fluxes from extra-galactic sources are more promising compared to single source fluxes in terms of event rates and because the signal may extend up to the highest energies. Nonetheless, these searches have to rely more on simulations than point source searches.
The basic concept is that the astrophysical signal should show up at high energies above the
atmospheric neutrino background given the harder spectrum. 
This implies that these searches are affected by theoretical uncertainties on high energy atmospheric 
neutrino fluxes (see Sec.~\ref{sub-neu}) and on the experimental errors in the high energy region were the statistics is low. Results at this conference from IceCube are preliminary for the muon topology \cite{ic22_kotoyo} and for cascades \cite{ic22_joanna}. Other results from Baikal \cite{baikal} and AMANDA \cite{amanda_uhe} are also shown in Fig.~\ref{right_fig6}. In the same figure, at EHE energies, current limits for cosmogenic neutrinos are also shown. These are produced by interactions of protons or nuclei with energies above the $\Delta$ production threshold of $\sim 10^{19.5}$ eV with the cosmic microwave background. IceCube presented results for the 22 string configuration \cite{ic22_ehe}. Also extensive air shower experiments look for such high energy neutrinos at the horizon \cite{auger,iori}, where the cosmic ray showers are suppressed by the large atmosphere layer. Radio detectors look for the coherent Cherenkov radiation produced by excess of electrons in neutrino induced showers in dense media (Askaryan effect) such as ice \cite{anita} and the lunar regolith \cite{atca,alvarez}. The main issue on the lunar regolith technique is the high threshold, above the expected peak of cosmogenic neutrino fluxes.

\section{New experiments and Instrumentation}
\label{sec6}

Various proposed experiments for gamma-astronomy have been presented at the conference, as well as upgrades to existing detectors. VERITAS is planning the relocation of one of the telescopes to improve the sensitivity by about 15\%, as well as upgrades of the cameras with high QE PMTs. If a fifth telescope would be added, the time to get to 1\% of Crab would be halved compared to the current one of $\sim 50$~hrs \cite{veritas2}. By the end of 2009 the 4 telescope system HESS will be upgraded with a new 28m-diameter telescope to lower the threshold to about 30 GeV \cite{hess2} and achieve an angular resolution before cuts of $\sim 0.25^{\circ}$ at $> 200$~GeV. MAGIC II has concluded the commissioning phase and has seen the first light. It is expected that 1\% of the Crab will be observed in 50 hrs for $E_{\gamma} \gtrsim 100$ GeV. The energy threshold is about 60~GeV \cite{magic2}. MAGIC is planning a  camera upgrade using hybrid photon detectors (HPDs) \cite{hpd}. These Hamamtsu devices consist of GaAsP photocathodes and a 3 mm diameter avalanche photodiode acting as an electron bombarded anode with additional internal gain. They have good single photoelectron resolution and achieve a quantum efficiency (QE) of about 50\% at 500 nm, though the QE at 300-400~nm decreases sharply. This aspect could make Geiger-mode avalanche photodiodes (G-APDs, previously known as SiPM) more interesting though wavelength shifters can be used for applications in gamma-astronomy. G-APDs devices are compact and light semiconductor photodetectors operated at very low voltage ($\lesssim 100$ V), that are not affected by light or magnetic fields, contrary to PMTs
\cite{gapd}. They are operated in Geiger discharge mode and the avalanched stopped using a quenching resistor or an active quenching circuit. The gain is of the order of $10^5 \div 10^6$ and linear in the overvoltage (bias voltage minus breakdown voltage) but since the breakdown voltage depends on temperature the bias voltage has to be adjusted when the temperature changes to keep the gain constant. This is the main challenge of operating these devices. Nowadays, they are also operated at room temperature and in the presence of high night sky background. A very interesting feature is the extremely high QE up to 65\% at 400~nm and 30\% at 300~nm. A prototype camera is operated at ETH in Zurich and they will be employed in the long-term monitoring of blazars project DWARF \cite{dwarf}.
R\&D on these new devices for IACTs cameras and other applications (e.g. direct detection of DM)
should be encouraged. Aside from improving camera detectors, other factors in sensitivity can be
gained by implementing at trigger level gamma/hadron discrimination techniques \cite{topo} or 
improving analyses tools \cite{becherini}.

\begin{figure*}[!t]
   \centerline{\subfloat[Summary of sensitivities to point-like sources of past, present and future IACTs and ground based gamma-astronomy detectors multiplied by energy (from \protect\cite{TibetArray}) for 1 year of observation for EAS ad 50 hrs for IACTs. The flux corresponds to $5\sigma$ or 10 event detection. The crosses correspond to the gmma-ray flux of the Crab Nebula observed by HEGRA.]
   {\includegraphics[width=3.in,height=2.5in]{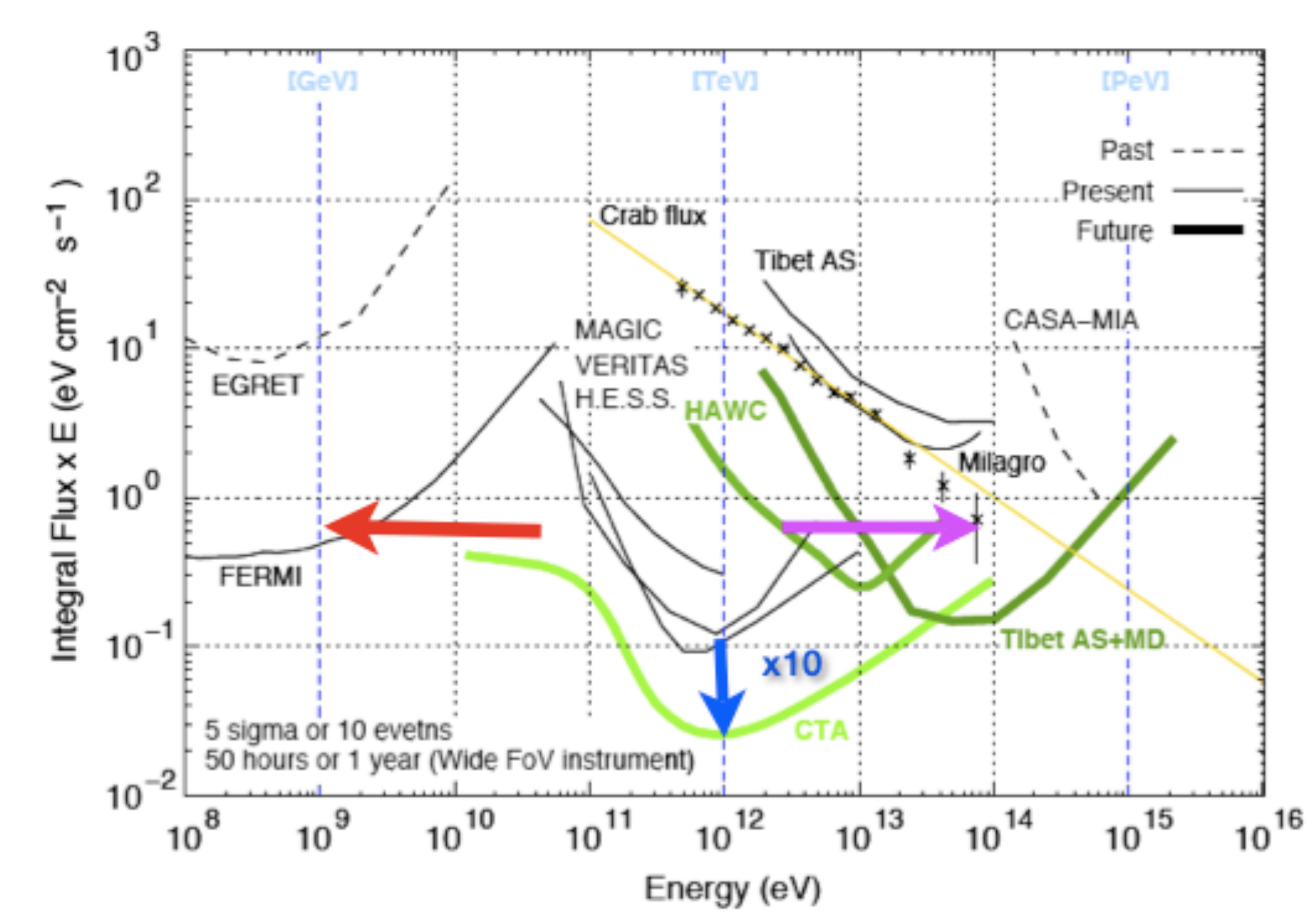} \label{left_fig7}}
              \hfil
              \subfloat[Angular resolution for AGIS with $0.1^{\circ}$ camera pixels vs energy for 36 telescopes at a reciprocal distance of 125 m\protect\cite{AGIS} compared to VERITAS \protect\cite{VERITAS_res} and the theoretical limit derived in Ref.~\protect\cite{Hoffmann}.]{\includegraphics[width=3.in,height=2.5in]{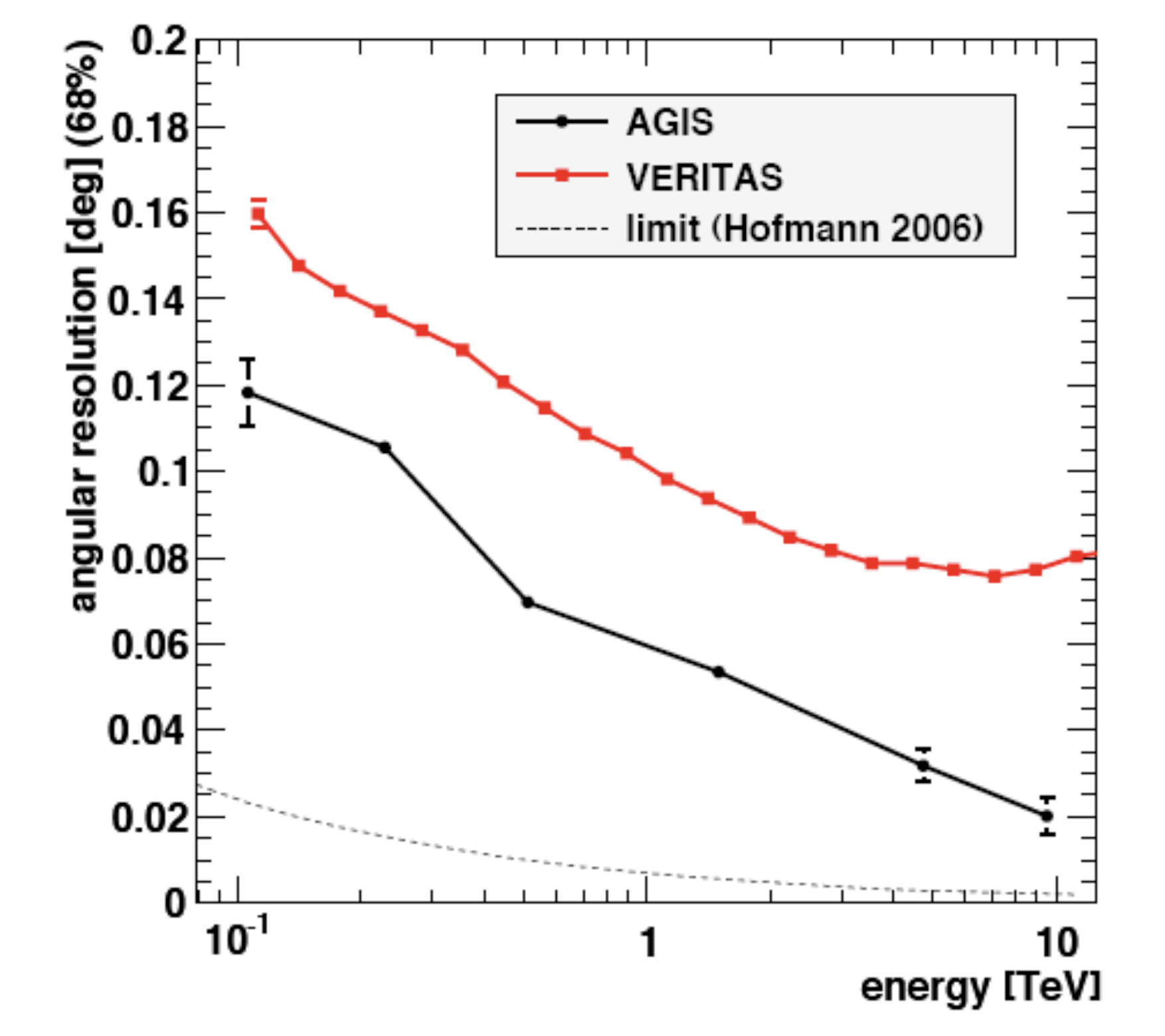} \label{right_fig7}}
             }
   \caption{}
   \label{fig7}
 \end{figure*}

A summary plot of sensitivities is in Fig.~\ref{left_fig7} for past, present and future projects.
The main aim in gamma astronomy is to fill the unexplored gap of energy between about 10 GeV where Fermi-LAT begins to run out of statistics and current IACT thresholds of a few hundreds of GeV. The goal is to gain about a factor of 10-20 in sensitivity by improving the angular resolution to pin down the morphology of sources and precisely determine the emission regions, increasing areas and duty cycle. It is also important to increase the field of view of detectors to observe  extended regions. 
In fact many observations of galactic sources, have revealed extended regions accelerating particles above tens of TeV, such as the Milagro Pevatrons \cite{pevatrons}, between which there is Geminga, never observed by IACTs, probably because it is too extended. These can be sources of galactic cosmic rays as well as it is possible that cosmic rays are produced in enormous super-bubbles in the interstellar medium \cite{butt} as some observations of extended hadronic sources seem to hint at \cite{milagro_regions,argo,ic22_anisotropy,tibet}. 
The scientific community is aiming at satisfying these goals by proposing arrays of tens of IACTs, such as CTA \cite{CTA} or AGIS \cite{AGIS}, and much cheaper extensive air shower (EAS) arrays, such as HAWC \cite{HAWC}. Other proposed arrays combine all of these techniques \cite{LHAASO}. The arrays of IACTs will guarantee a better angular resolution by at least a factor of two with respect to current detectors (see Fig.~\ref{right_fig7}). CTA is following a more conservative approach for what concern telescope mirrors than AGIS that is proposing new Schwartzchild-Couder optics with a primary and secondary mirror.
CTA is planning an array with variable size telescopes and variable baselines: a core of about 4-6 
large telescopes with diameters $\sim 20-30$ m and FoV $\sim 3^{\circ}-4^{\circ}$ will be dedicated to low energies up to about 10 GeV, an array of Davies-Cotton telescopes with 10-12 m diameters and FoV of $6^{\circ}-8^{\circ}$ for intermediate energies and many $\sim 6$~m diameter telescopes with large FoV ($8^{\circ}-10^{\circ}$) at large baselines for covering large energies. Optimal sites for such arrays should be between 2000-3000 a.s.l. Hopefully it will be possible to build two arrays in the 2 hemispheres to cover the entire sky by joining international efforts.
A conceptual design of a 23 m diameter telescope capable of achieving a low energy threshold, with low weight (50 tons) and with a carbon fiber mechanical structure for fast rotations optimal for transient sources was presented in \cite{eckart}.

The idea behind HAWC \cite{HAWC} is to have an upgraded detector respect to Milagro, with a modular structure, that could be ready to operate in less than 3 years to monitor the sky at TeV energies, whereas Fermi would cover the lower energy range. With a modular structure made of 300 tanks of water spread over an area of about $150 \times 150$ m$^2$ it is possible to improve the hadron/gamma discrimination with respect to Milagro. By building the detector at 4100 m$^2$ on a Sierra Negra (Mexico) plateau, the energy threshold can be lowered thanks to the higher statistics of particles close to shower maximum. Though HAWC will monitor the available portion of the sky at Mexican latitudes with 100\% of duty cycle, the angular resolution (not yet optimized) will be worse (about $0.25^{\circ}$ for $E_{\gamma} \gtrsim 5$~TeV) by about an order of magnitude than CTA/AGIS (see Fig.~\ref{right_fig7}). This is comprehensible since the cost is more than an order of magnitude lower for HAWC with respect to CTA/AGIS. This complementary strategy is a good approach to address the remaining discovery areas for gamma-astronomy, such as transient sources and gamma-ray bursts, and long term monitoring of blazars and extended regions of hadronic production.

In the field of neutrino astronomy R\&D programs are also going on.
The KM3NeT consortium has finalized the conceptual design of a future cubic-kilometer scale 
detector in the Mediterranean with a sensitivity 2-3 times better at high energy than IceCube thanks
to the better angular resolution achievable in sea water \cite{km3net}. The project is studying 
tower structures and their configuration to maximize performances. Since this project is sensitive to 
a wide region of interest for galactic sources and to the Galactic Centre, the configuration of strings
should be such that at 1-100 TeV the sensitivity of the detector is better than IceCube sensitivity.
Finally, together with radio programs to extend neutrino astronomy to the scale of 100 km$^2$,
R\&D on acoustic detection of UHE neutrinos is being performed at the ANTARES \cite{amadeus} and Baikal \cite{baikal_acoustic} sites in water and at IceCube site in Antarctica \cite{spats}. Acoustic detection of neutrinos is based on the thermal energy deposition of the induced particle cascade that produces a 
bipolar acoustic pulse in the kHz range. Acoustic energy thresholds for neutrinos are around $10^{18} \div 10^{19}$~eV.
To extend NTs to 100 km$^2$, an attenuation length much larger than the one in the optical is desirable for new techniques. Water seems more promising than ice since the attenuation length
is of the order 1 km, hence about a factor of 10 respect to the optical one. In ice it seems to be of the order of 200-350 m. Given the large scattering of light in ice, this is about a factor of 10 with respect to the attenuation length in ice, but still much smaller than in sea water. On the other hand, water is a noisy environment. In sea water mammals are detected as well as surface noise correlated to winds, while in Baikal noise increases during ice coverage/melting periods. A prototype \cite{baikal_acoustic} made of an acoustic antenna 150 m deep and an imitator has been installed close to NT200+. The achieved angular resolution is $1.5^{\circ}$ in azimuth and $0.5^{\circ}$ in zenith. About 7000 acoustic pulses were detected between Apr.-May 2009 and all of them are down-going but one
that is up-going and is not explained. Hence, this technique is not background free, and background understanding is challenging.
AMADEUS (ANTARES Modules for Acoustic DEtection Under the Sea) \cite{amadeus} is using 3 storeys of the ANTARES detector to host acoustic devices with a bandpass filter from 1 kHz to 100 kHz. They proved capability of identification of transient backgrounds with the beamforming technique, that is the coherent sum of sound waves sampled by hydrophones, and produced a map where most of the noise is from the upper hemisphere and in the lower hemisphere the ANTARES pingers on the lines are visible.
\cite{baikal_acoustic}

\section{Summary and Outlook}
\label{sec7}
 
Most of the results presented at this conference on neutrino astronomy and DM are upper limits, even if IceCube has showed point-source results for 6 months of data taken with 1/2 of its full configuration (40 strings).
These limits are beginning to be in the region of interest between $E^2 \frac{dN}{dE} \sim 10^{-12}-10^{-11}$ TeV cm$^{-2}$ s$^{-1}$ for galactic sources and diffuse flux limits are approaching the Waxman \& Bahcall upper limit on extra-galactic neutrino fluxes.  According to estimates based on gamma observations, it is possible that $\sim 5$ years of IceCube will be needed to achieve $5\sigma$ significance for Milagro Pevatrons \cite{halzen_pevatrons}.
Another detector in the opposite hemisphere would cover the entire sky and be sensitive to the Galactic Centre region, while IceCube has sensitivity to Cygnus that is observed by Milagro up to tens of TeV.
Studies and R\&D for detectors of the order of 10-100 times IceCube are a promising strategy for the future, but no technique is yet available at a reasonable cost for the same energy threshold as current NTs. In fact, radio techniques that are very promising to achieve 100 km$^2$, have thresholds of the order of $10^{18}$ eV suitable for cosmogenic neutrinos but considerably higher than IceCube.

Gamma astronomy future programs aim at pushing the energy threshold as low as possible
to fill the gap from Fermi-GLAST and possibly observe close-by gamma-ray bursts with ground based arrays in the TeV range.
In the future, arrays of IACT telescopes can be suitable for this scope joining the European and American communities. R\&D for camera detectors should be strongly encouraged. In the meanwhile large FoV arrays with 100\% duty cycles could monitor the TeV sky complementing at higher energies Fermi-GLAST.

\section{Acknowledgments}
I would like to thank F. Arneodo, F. Ronga, F. Halzen and L. Anchordoqui for sending me comments on the manuscript and reading it all! I also would like to thank G. Wikstrom and Jim Braun for providing me the script to 
modify Fig.~\ref{fig3}.


\begin{thebibliography}{199}
\bibitem{dm_antares} G.~M.~A.~Lim for the ANTARES Collaboration, \emph{First results on the search for dark matter in the Sun with the ANTARES neutrino telescope}, HE 2.3 0031.
\end{thebibliography}
\end{document}